\documentclass[11pt]{article}
\usepackage{fullpage}
\usepackage[utf8]{inputenc}
\usepackage[english]{babel}
\usepackage{amsmath, amsfonts, amssymb, amsthm}
\usepackage{url}

\usepackage{tikz}
\usetikzlibrary{calc,arrows,shapes,backgrounds,patterns,fit,decorations,decorations.pathmorphing}

\usepackage[ruled]{algorithm}
\usepackage[noend]{algpseudocode}

\DeclareMathOperator{\overlap}{overlap}
\DeclareMathOperator{\pref}{pref}
\DeclareMathOperator{\suff}{suff}

\DeclareMathOperator{\poly}{poly}
\DeclareMathOperator{\inp}{input}
\DeclareMathOperator{\indegree}{indegree}
\DeclareMathOperator{\outdegree}{outdegree}

\renewcommand{\geq}{\geqslant}
\renewcommand{\leq}{\leqslant}

\newcommand{\ab}[1]{#1}

\tikzstyle{hgedge}=[->,gray!40!white]
\tikzstyle{anypath}=[->,dashed]
\tikzstyle{vertex}=[draw,ellipse,inner sep=0.5mm,minimum size=4mm]
\tikzstyle{inputvertex}=[draw,rectangle,inner sep=.5mm,minimum size=5mm]

\newenvironment{mypic}{\begin{center}\begin{tikzpicture}[>=latex,line width=.3mm,scale=0.8,transform shape]}{\end{tikzpicture}\end{center}}

\newtheorem{lemma}{Lemma}
\newtheorem{theorem}{Theorem}
\newtheorem{proposition}{Proposition}
\newtheorem{claim}{Claim}
\newtheorem{corollary}{Corollary}

\begin{document}

\sloppy
\begin{titlepage}
\title{Collapsing Superstring Conjecture}
\author{
Alexander Golovnev\thanks{Harvard University. Supported by a Rabin Postdoctoral Fellowship.}
\and
Alexander~S.~Kulikov\thanks{Steklov Institute of Mathematics at St.~Petersburg, Russian Academy of Sciences}
\and
Alexander Logunov\footnotemark[2]
\and
Ivan Mihajlin\thanks{University of California, San Diego}
\and 
Maksim Nikolaev\thanks{St.~Petersburg State University}
}
\maketitle
\thispagestyle{empty}

\begin{abstract}
In the Shortest Common Superstring (SCS) problem, one is given a collection of strings, and needs to find a shortest string containing each of them as a~substring. SCS admits $2\frac{11}{23}$-approximation in polynomial time (Mucha, SODA'13). While this algorithm and its analysis are technically involved, the $30$ years old Greedy Conjecture claims that the trivial and efficient Greedy Algorithm gives a~$2$-approximation for SCS. 

We develop a graph-theoretic framework for studying approximation algorithms for SCS. The framework is reminiscent of the classical 2-approximation for Traveling Salesman: take two copies of an optimal solution, apply a trivial edge-collapsing procedure, and get an approximate solution. In this framework, we observe two surprising properties of SCS solutions, and we conjecture that they hold for all input instances. 
The first conjecture, that we call Collapsing Superstring conjecture, claims that there is an elementary way to transform any solution repeated twice into the same graph~$G$. This conjecture would give an elementary 2-approximate algorithm for SCS. The second conjecture claims that not only the resulting graph~$G$ is the same for all solutions, but that~$G$ can be computed by an elementary greedy procedure called Greedy Hierarchical Algorithm.

While the second conjecture clearly implies the first one, perhaps surprisingly we prove their equivalence. We support these equivalent conjectures by giving a proof for the special case where all input strings have length at most~$3$ (which until recently had been the only case where the Greedy Conjecture was proven). We also tested our conjectures on millions of instances of SCS.

We prove that the standard Greedy Conjecture implies Greedy Hierarchical Conjecture, while the latter is sufficient for an efficient greedy 2-approximate approximation of SCS. Except for its (conjectured) good approximation ratio, the Greedy Hierarchical Algorithm provably finds a $3.5$-approximation, and finds \emph{exact} solutions for the special cases where we know polynomial time (not greedy) exact algorithms: (1)~when the input strings form a~spectrum of a~string (2)~when all input strings have length at most~$2$.
\end{abstract}

\end{titlepage}

\section{Introduction}
\label{sec:intro}
The {\em shortest common superstring problem} (abbreviated as SCS) is:
given a~set of strings, find a~shortest string that contains all of them as
substrings. This problem finds applications in genome assembly~\cite{waterman1995introduction, pevzner2001eulerian}, and data compression~\cite{GMS1980, phdthesis, storer1987data}. We refer the reader to the excellent surveys~\cite{gevezes2014shortest, mucha2007tutorial} for an overview of SCS, its applications and algorithms.  SCS is known to be $\mathbf{NP}$-hard~\cite{GMS1980} and even $\mathbf{MAX}$-$\mathbf{SNP}$-hard~\cite{BJLTY1991}, but it admits constant-factor approximation in polynomial time.

The best known approximation ratios are $2\frac{11}{23}$ due to Mucha~\cite{M2013} 
and $2\frac{11}{30}$ due to Paluch~\cite{P14} 
(see \cite[Section~2.1]{GKM13} for an overview of the 
previous approximation algorithms
and inapproximability results). While these approximation algorithms use an algorithm for Maximum Weight Perfect Matching as a~subroutine, the $30$ years old \emph{Greedy Conjecture}~\cite{storer1987data, TU1988, T1989, BJLTY1991} claims that the trivial \emph{Greedy Algorithm}, whose pseudocode is given in Algorithm~\ref{algo:ga}, is 2-approximate. Ukkonen~\cite{ukkonen1990linear} shows that for a fixed alphabet, the Greedy Algorithm can be implemented in linear time. It should be noted that GA is not deterministic as we do not specify how to break ties in case when there are many pairs of strings with maximum overlap. For this reason, GA may produce different superstrings for the same input.

\begin{algorithm}[ht]
\caption{Greedy Algorithm (GA)}\label{algo:ga}
\hspace*{\algorithmicindent} \textbf{Input:} set of strings~${\cal S}$.\\
\hspace*{\algorithmicindent} \textbf{Output:} a~superstring~for $\mathcal{S}$.
\begin{algorithmic}[1]
\While{$\mathcal{S}$ contains at least two strings}
\State extract from $\mathcal{S}$ two strings with the maximum overlap
\State add to $\mathcal{S}$ the shortest superstring of these two strings
\EndWhile
\State return the only string from $\mathcal{S}$
\end{algorithmic}
\end{algorithm}

\newtheorem*{gc}{Greedy Conjecture}
\begin{gc}
For any set of strings~$\mathcal{S}$, $\operatorname{GA}(\mathcal{S})$ constructs a~superstring that is at most twice longer than an optimal one.
\end{gc}

Blum et al.~\cite{BJLTY1991} prove that the Greedy Algorithm returns a $4$-approximation of SCS, and Kaplan and Shafrir~\cite{KS2005} improve this bound to~$3.5$. A~slight modification of the Greedy Algorithm gives a $3$-approximation of SCS~\cite{BJLTY1991}, and other greedy algorithms are studied from theoretical~\cite{BJLTY1991,rivals2018superstrings} and practical perspectives~\cite{romero2004experimental, cazaux2018practical}.

It is known that the Greedy Conjecture holds for the case when all input strings have length at most $3$~\cite{TU1988, cazaux20143}, and it was recently shown to hold in the case of strings of length $4$~\cite{kulikov2015greedy}. Also, the Greedy Conjecture holds if the Greedy Algorithm happens to merge strings in a particular order~\cite{weinard2006greedy, laube2005conditional}. The Greedy Algorithm gives a $2$-approximation of a~different metric called compression~\cite{TU1988}. The compression is defined as the sum of the lengths of all input strings minus the length of a superstring
(hence, it is the number of symbols saved with respect to a~naive superstring resulting from concatenating the input strings).


Most of the approaches for approximating SCS are based on the
{\em overlap graph} or the equivalent \emph{suffix graph}. The suffix graph has input strings as nodes, and a pair of nodes 
is joined by an~arc of weight equal to their suffix (see Section~\ref{sec:def_scs} for formal definitions of overlap and suffix).
SCS is equivalent to (the asymmetric version of) the Traveling Salesman Problem (TSP) in the suffix graph. While TSP cannot be approximated within any polynomial time computable function unless $\mathbf{P}=\mathbf{NP}$~\cite{SG1976}, its special case corresponding to SCS can be approximated within a constant factor.\footnote{We remark that SCS is also a special case of TSP for costs satisfying the triangle inequaliy. This case of TSP can be approximated within a constant factor~\cite{svensson2018constant}, but this factor is currently much worse than that for SCS.} We do not know the full characterization of the graphs in this special case, but we know some of their properties: Monge inequality~\cite{monge} and Triple inequality~\cite{weinard2006greedy}. These properties are provably not sufficient for proving Greedy Conjecture~\cite{weinard2006greedy, laube2005conditional}. 

While the overlap and suffix graphs give a~convenient graph structure, our current knowledge of their properties is provably not sufficient for showing strong approximation factors. Thus, the known approximation algorithms (including the Greedy Algorithm) estimate the approximation ratio via the overlap graph, and also separately take into account some string properties not represented by the overlap graph. The goal of this work is to develop a simple combinatorial framework which captures all features of the input strings needed for proving approximation ratios of algorithms.

\subsection{Our contributions}

We continue the study of the so-called {\em hierarchical graph}
introduced by Golovnev et al.~\cite{scs_exact}. (See also~\cite{cr16} for a related notion of the superstring graph.) This graph is designed specifically 
for the SCS problem, in some sense it generalizes de~Bruijn graph, and it contains more information about the input strings
than just all pairwise overlaps. Given an instance of SCS, the vertex set of the corresponding hierarchical graph is just the set of substrings of all the input strings. For a string $s$ and two symbols $\alpha, \beta$, the graph contains the arcs: $(s,s\alpha)$ and $(\beta s, s)$. Now, every superstring of the given set of string corresponds to an Eulerian walk in the hierarchical graph (which passes through the vertices corresponding to the input strings), and vice versa. (See Section~\ref{sec:def_hier} for formal definition and statements.)

\subsubsection{Collapsing Conjecture}
We define a simple normalization procedure of a walk in the hierarchical graph: replace the pair of arcs $(\alpha s, \alpha s \beta), (\alpha s \beta, s\beta)$ with the pair $(\alpha s, s ), (s , s\beta)$ as long as it does not violate connectivity of the walk. It is easy to see that such a normalization never increases the length of the corresponding solution of SCS. 
First, we observe a surprising property of this normalization procedure: if one takes \emph{any} solution, doubles all of its arcs in the hierarchical graph, and then applies the normalization procedure, then the resulting set of arcs is always the same (i.e., it does not depend on the initial solution). Collapsing Conjecture makes this observation formal (see Section~\ref{sec:collapsing}). Note that this conjecture implies an extremely simple 2-approximate algorithm for Shortest Common Superstring: take any solution (for example, write down all input strings one after another), then double each arc in the hierarchical graph, and apply the simple normalization procedure. This procedure will result in some superstring $S$. On the other hand, if one started with an optimal solution, doubled each of its arcs, and normalized the result, then the resulting solution would have length at most twice the length of the optimal solution. By Collapsing Conjecture, this resulting superstring would also be $S$, which implies that $S$ is a $2$-approximation.

\subsubsection{Greedy Hierarchical Conjecture}
We also propose a simple and natural greedy algorithm for SCS in the hierarchical graph: start from the nodes corresponding to the input strings, and greedily build an Eulerian walk passing through all of them. 
While this Greedy Hierarchical Algorithm (GHA) is as simple as the Greedy Algorithm (GA), it provably performs better in some cases. For example, there are two well-known polynomially solvable special cases of SCS: strings of length~$2$ and
a~spectrum of a~string. While GA does not always find optimal solutions in these cases, GHA solves them exactly (see Sections~\ref{sec:ghatwo} and~\ref{sec:ghaspectrum}).

Greedy Hierarchical Conjecture (see Section~\ref{sec:greedy_hier}) claims that the set of arcs produced by GHA exactly matches the set of arcs from the Collapsing Conjecture: whichever initial solution one takes, after doubling its arcs and normalization, the resulting set of arcs is exactly the solution found by GHA. Clearly, this conjecture implies Collapsing Conjecture. Perhaps surprisingly, we prove that the two conjectures are equivalent (see Section~\ref{sec:equiv}): if all doubled solutions after normalization result in the same set of arcs, then this set of arcs is the GHA solution.

The weak form of Greedy Hierarchical Conjecture claims that GHA is a 2-approximate algorithm for SCS. 
We prove (see Section~\ref{sec:gr_im_wghc}) that GHA is an instantiation of GA with some tie-breaking rule. That is, there is an algorithm which always merges some pair of strings with the longest overlap and outputs the same solution as GHA. This result has two consequences. First, by the known results for GA, we immediately have that GHA finds a $3.5$-approximation for SCS. Second, this gives us that Greedy Conjecture implies Weak Greedy Hierarchical Conjecture. 

\subsubsection{Evidence for the Conjectures}
We support the Collapsing Conjecture (and the equivalent Greedy Hierarchical Conjecture) by proving its special case and verifying it empirically. We prove the conjecture for the special case where all input strings have length at most $3$, which until recently had been the only case where the Greedy Conjecture was proven (see Section~\ref{subsec:scs3}). Despite tesing the conjecture on millions of datasets (both hand-crafted and generated randomly according to various distributions), we have not found a counter-example. Note that even the Weak Greedy Hierarchical Conjecture suffices for getting a $2$-approximation for SCS, and this conjecture is not harder to prove than the standard Greedy Conjecture.
We implemented the Greedy Hierarchical Algorithm~\cite{github}, and we invite the reader to its~web interface~\cite{webpage} to see step by step executions of the described algorithms and to verify the conjectures on
custom datasets.

\section{Definitions}
\subsection{Shortest Common Superstring Problem}
\label{sec:def_scs}
\begin{figure}[ht]
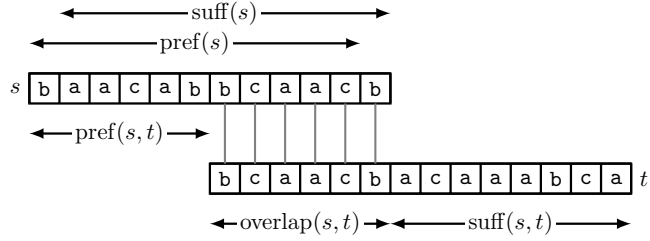

\begin{mypic}
\begin{scope}

\draw (0,2) rectangle (6,2.5);
\draw[step=5mm] (0,2) grid (6,2.5);
\node[left] at (0,2.25) {$s$};
\draw (3,0.5) rectangle (10,1);
\draw[step=5mm] (3,0.5) grid (10,1);
\node[right] at (10,0.75) {$t$};

\foreach \x in {3.25, 3.75, ..., 5.75}
  \draw[gray,thick] (\x,2) -- (\x,1);

\foreach \f/\t/\y/\lab in {0/3/1.5/{\pref(s,t)}, 
3/6/0/{\overlap(s,t)}, 6/10/0/{\suff(s,t)}, 0/5.5/3/\pref(s),
0.5/6/3.5/\suff(s)}
  \path (\f,\y) edge[<->] node[rectangle,inner sep=0.5mm,fill=white] {\strut $\lab$} (\t,\y);
  
\foreach \x/\a in {0/b, 0.5/a, 1/a, 1.5/c, 2/a, 2.5/b, 3/b, 3.5/c, 4/a, 4.5/a, 5/c, 5.5/b}
  \node at (\x+0.25,2.25) {\tt \a};
\foreach \x/\a in {3/b, 3.5/c, 4/a, 4.5/a, 5/c, 5.5/b, 6/a, 6.5/c, 7/a, 7.5/a, 8/a, 8.5/b, 9/c, 9.5/a}
  \node at (\x+0.25,0.75) {\tt \a};
\end{scope}
\end{mypic}
\caption{Pictorial explanations of $\pref$, $\suff$, and $\overlap$ functions.}
\label{fig:overlap}
\end{figure}

For a string $s$, by $|s|$ we denote the length of $s$. 
For strings~$s$ and~$t$, by $\overlap(s,t)$
we denote the longest suffix of~$s$ that is also 
a~prefix of~$t$. By $\pref(s,t)$
we denote the first $|s|-|\overlap(s,t)|$ symbols of $s$.
Similarly, $\suff(s,t)$ is the last
$|t|-|\overlap(s,t)|$ symbols of~$t$. 
By $\pref(s)$ and $\suff(s)$ we denote, respectively,
the first and the last $|s|-1$ symbols of~$s$. See Figure~\ref{fig:overlap} for a~visual explanation. We denote the empty string by $\varepsilon$.

\begin{figure}[ht]
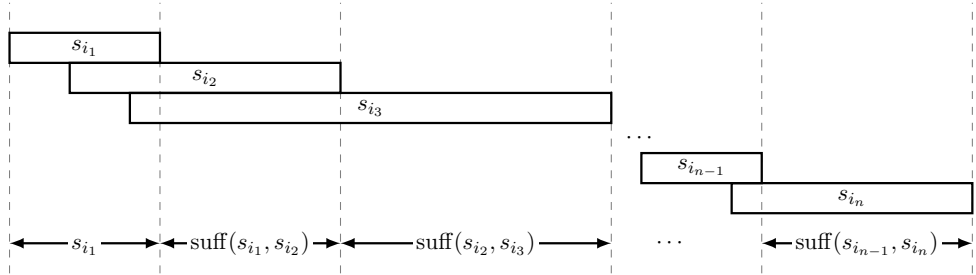

\begin{mypic}

\foreach \x in {0, 2.5, 5.5, 10, 12.5, 16}
  \draw[dashed,gray,thin] (\x,-0.5) -- (\x,4);

\foreach \x/\y/\len/\label in {0/3/5/s_{i_1}, 1/2.5/9/s_{i_2}, 2/2/16/s_{i_3}, 10.5/1/4/s_{i_{n-1}}, 12/0.5/8/s_{i_n}} {
  \draw (\x,\y) rectangle (\x+0.5*\len,\y+0.5);
  \node at (\x+0.25*\len,\y+0.25) {$\label$};
}

\foreach \f/\t/\label in {0/2.5/{s_{i_1}}, 2.5/5.5/{\suff(s_{i_1}, s_{i_2})}, 5.5/10/{\suff(s_{i_2}, s_{i_3})}, 12.5/16/{\suff(s_{i_{n-1}}, s_{i_n})}}
  \path (\f,0) edge[<->] node[rectangle,inner sep=0.5mm,fill=white] {\strut $\label$} (\t,0);

\node at (11,0) {$\dotsb$};
\node at (10.5,1.75) {$\dotsb$};
\end{mypic}
\caption{SCS is a~permutation problem. The length of a~superstring corresponding to a~permutation $(s_{i_1}, \dotsc, s_{i_n})$ is $|s_{i_1}|$ plus the sum of the lengths of suffixes of consecutive pairs of strings. It is also equal to $\sum_{i=1}^n |s_i|-\sum_{j=1}^{n-1}|\overlap(s_{i_j}, s_{i_{j+1}})|$.}
\label{fig:permutation}
\end{figure}

Throughout the paper by ${\cal S}=\{s_1, \dots, s_n\}$ we denote
the set of~$n$ input strings. We assume that no input string is a~substring of another (such a~substring can be removed from $\mathcal{S}$ in the preprocessing stage). Note that SCS is a~{\em permutation problem}: to find a~shortest string containing all $s_i$'s in a~{\em given order} one just
overlaps the strings in this order, see Figure~\ref{fig:permutation}. (This simple observation relates SCS to other permutation problems, including various versions of the Traveling Salesman Problem.) It will prove convenient to view the SCS problem as a~problem of finding an optimum permutation.
It should be noted at the same time that the correspondence between permutations and superstrings is not one-to-one: there are superstrings that do not correspond to any permutation. For example, the concatenation of input strings is clearly a~superstring, but it ignores the fact that neighbor strings may have non-trivial overlaps and for this reason may fail to correspond to a~permutation. Still, clearly, any {\em shortest} superstring corresponds to a~permutation of the input strtings.

\subsection{Hierarchical Graph}
\label{sec:def_hier}
For a~set of strings~${\cal S}$, the~\emph{hierarchical graph} $HG=(V,E)$ is a~weighted directed graph with $V=\{v \colon \text{$v$ is a~substring of some $s \in {\cal S}$}\}$. For every $v \in V,\, v \neq \varepsilon$, the set of arcs~$E$ contains an {\em up-arc} $(\pref(v), v)$ of weight~1 and a {\em down-arc} $(v, \suff(v))$ of weight~0. The meaning of an up-arc is appending one symbol to the end of the current string (and that is why it has weight~1), whereas the meaning of a down-arc is cutting down one symbol from the beginning of the current string.
%
Figure~\ref{fig:hgex}(a) gives an example of the hierarchical graph and shows that the terminology of up- and down-arcs comes from placing all the strings of the same length at the same level, where the $i$-th level contains strings of length~$i$.  In all the figures in this paper, the input strings are shown in rectangles, while all other vertices are ellipses.

\newcommand{\we}[4]{
\begin{scope}[xshift=#1mm,yshift=#2mm]
\foreach \n/\x/\y in {aaa/0/3, cae/1/3, aec/3/3, eee/4/3}
  \node[inputvertex] (\n) at (\x,\y) {\tt \n};
\foreach \n/\x/\y in {aa/0/2, ca/1/2, ae/2/2, ec/3/2, ee/4/2, a/1/1, c/2/1, e/3/1}
  \node[vertex] (\n) at (\x,\y) {\tt \n};
\node[vertex] (eps) at (2,0) {$\varepsilon$};
\foreach \f/\t/\a in {eps/e/10, e/eps/10, eps/c/10, c/eps/10, eps/a/10, a/eps/10, a/aa/10, aa/a/10, aa/aaa/10, aaa/aa/10, c/ca/0, ca/cae/0, cae/ae/0, ae/aec/0, aec/ec/0, ee/eee/10, eee/ee/10, e/ee/10, ee/e/10, ca/a/0, a/ae/0, ae/e/0, e/ec/0, ec/c/0}
  \path (\f) edge[hgedge,bend left=\a] (\t);
  
\node at (2,-1) {(#3)};

#4
\end{scope}
}

\begin{figure}[!ht]
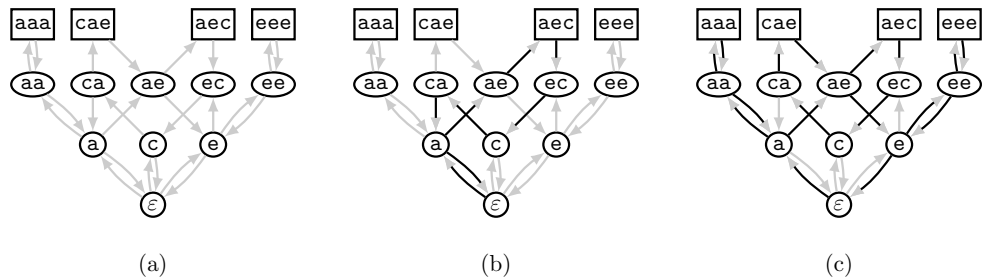

\begin{mypic}
\we{0}{0}{a}{}

\we{57}{0}{b}{
\foreach \f/\t/\a in {eps/a/10, a/ae/0, ae/aec/0, aec/ec/0, ec/c/0, c/ca/0, ca/a/0, a/eps/10}
  \path (\f) edge[hgedge,bend left=\a,draw=black,thick] (\t);
}

\we{114}{0}{c}{
\foreach \f/\t/\a in {eps/a/10, a/aa/10, aa/aaa/10, aaa/aa/10, aa/a/10, a/ae/0, ae/aec/0, aec/ec/0, ec/c/0, c/ca/0, ca/cae/0, cae/ae/0, ae/e/0, e/ee/10, ee/eee/10, eee/ee/10, ee/e/10, e/eps/10}
  \path (\f) edge[hgedge,bend left=\a,draw=black,thick] (\t);
}
\end{mypic}
\caption{(a)~Hierarchical graph for the~dataset $\mathcal{S}=\{{\tt aaa}, {\tt cae}, {\tt aec}, {\tt eee}\}$. (b)~The~walk $\varepsilon \to {\tt a} \to {\tt ae} \to {\tt aec} \to {\tt ec} \to {\tt c} \to {\tt ca} \to {\tt a} \to \varepsilon$ has length (or weight)~4 and spells the~string {\tt aeca} of length~4. (c)~An~optimal superstring for~$\mathcal{S}$ is {\tt aaaecaeee}. It has length~9, corresponds to the~permutation $({\tt aaa}, {\tt aec}, {\tt cae}, {\tt eee})$, and defines the~walk of length~9 shown in black.}
\label{fig:hgex}
\end{figure}

What we are looking for in this graph is a shortest
walk from $\varepsilon$ to $\varepsilon$ going
through all the nodes from~$\mathcal{S}$.
It is not difficult to see that the length of a~walk 
from $\varepsilon$ to $\varepsilon$ equals the 
length of the string spelled by this walk. 
This is just because each up-arc has 
weight~$1$ and adds one symbol to 
the current string. See Figure~\ref{fig:hgex}(b) for an~example.

Hence, the SCS problem is equivalent to finding a~shortest closed walk from $\varepsilon$ to $\varepsilon$ that visits all nodes from~${\cal S}$. Note that a walk may contain repeated nodes and arcs. The multiset of arcs of such a~walk must be Eulerian (each vertex must have the same in- and out-degree, and the set of arcs must be connected). It will prove convenient to define an~{\em Eulerian solution} in a~hierarchical graph as an~Eulerian multiset of arcs~$D$ that goes through $\varepsilon$ and all nodes from~${\cal S}$. Given such a~solution~$D$, one can easily recover an Eulerian cycle (that might not be unique). This cycle spells a~superstring of~${\cal S}$ of the same length as~$D$. Figure~\ref{fig:hgex}(c) shows an~optimal Eulerian solution.

\begin{figure}[ht]
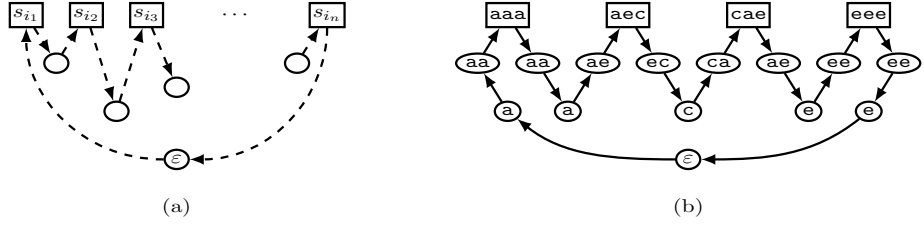

\begin{mypic}
\begin{scope}[yscale=0.8]
\foreach \n/\x/\y in {aaa/0/3, cae/4/3, aec/2/3, eee/6/3}
  \node[inputvertex] (\n) at (\x,\y) {\tt \n};

\foreach \n/\t/\x/\y in {aa1/aa/-0.5/2, aa2/aa/0.5/2, ae1/ae/1.5/2, ec/ec/2.5/2, ca/ca/3.5/2, ae2/ae/4.5/2, ee1/ee/5.5/2, ee2/ee/6.5/2, a1/a/0/1, a2/a/1/1, c/c/3/1, e1/e/5/1, e2/e/6/1}
  \node[vertex] (\n) at (\x,\y) {\tt \t};

\node[vertex] (eps) at (3,0) {$\varepsilon$};

\foreach \f/\t in {aa1/aaa, aaa/aa2, ae1/aec, aec/ec, ca/cae, cae/ae2, ee1/eee, eee/ee2, a1/aa1, aa2/a2, a2/ae1, ec/c, c/ca, ae2/e1, e1/ee1, ee2/e2}
  \draw[->] (\f) -- (\t);
  
\path (eps) edge[->,out=180,in=-45] (a1);
\path (e2) edge[->,out=-135,in=0] (eps);

\foreach \n/\x in {1/-8, 2/-7, 3/-6, n/-3}
  \node[inputvertex] (\n) at (\x,3) {$s_{i_{\n}}$}; 
  
\node at (-4.5,3) {$\dotsb$};

\foreach \n/\x/\y in {12/-7.5/2, 23/-6.5/1, 34/-5.5/1.5, 67/-3.5/2}
  \node[vertex] (\n) at (\x,\y) {}; 
  
\foreach \f/\t in {1/12, 12/2, 2/23, 23/3, 3/34, 67/n}
  \draw[->,anypath] (\f) -- (\t);
  
\node[vertex] (eps) at (-5.5,0) {$\varepsilon$};
\path (eps) edge[anypath,->,out=180,in=-90] (1);
\path (n) edge[anypath,->,out=-90,in=0] (eps);

\node at (-5.5,-1) {(a)};
\node at (3,-1) {(b)};
\end{scope}
\end{mypic}
\caption{(a)~A~schematic illustration of a~normalized Eulerian solution.
(b)~Untangled optimal Eulerian solution from Figure~\ref{fig:hgex}(c).}
\label{fig:hgperm}
\end{figure}

A solution to SCS defines a permutation $(s_{i_1}, \dotsc, s_{i_n})$ of the input strings, and this permutation naturally gives a~``zig-zag'' Eulerian solution in the hierarchical graph:
\begin{align}
\label{eq:zigzag}
\varepsilon \to s_{i_1} \to \overlap(s_{i_1}, s_{i_2}) \to s_{i_2} \to
\overlap(s_{i_2}, s_{i_3}) \to \dotsb 
\to s_{i_n} \to \varepsilon \, .
\end{align}
This Eulerian solution is shown schematically in Figure~\ref{fig:hgperm}(a). This schematic illustration is over simplified as the shown path usually has many self-intersections. Still, this point of view is helpful in understanding the algorithms presented later in the text. 
Figure~\ref{fig:hgperm}(b) shows an ``untangled'' optimal Eulerain solution from Figure~\ref{fig:hgex}(c): by contracting nodes with equal labels into the same node, one gets exactly the solution from Figure~\ref{fig:hgex}(c).

Not every Eulerian solution in the hierarchical graph has a~nice zig-zag
structure described above. In the next section, we introduce a~normalization procedure (that we call collapsing)
that allows us to focus on nice Eulerian solutions only. 

\subsection{Normalizing a~Solution}
\label{sec:def_normal}
\begin{figure}[ht]
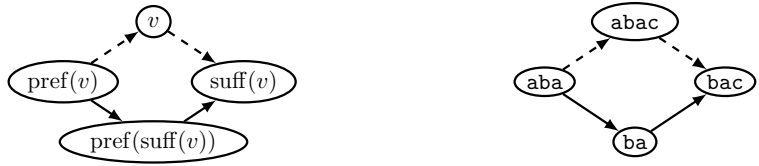

\begin{mypic}
\begin{scope}[minimum size=6mm]
\node[vertex] (a) at (0, 1.5) {$\pref(v)$};
\node[vertex,inner sep=1mm] (b) at (1.5, 2.5) {$v$};
\node[vertex] (c) at (3, 1.5) {$\suff(v)$};
\node[vertex] (d) at (1.5, 0.5) {$ \pref(\suff(v))$};
\draw[->,dashed] (a) -- (b);
\draw[->,dashed] (b) -- (c);
\draw[->] (a) -- (d);
\draw[->] (d) -- (c);

\begin{scope}[xshift=80mm]
\node[vertex] (a) at (0, 1.5) {\tt aba};
\node[vertex,inner sep=1mm] (b) at (1.5, 2.5) {\tt abac};
\node[vertex] (c) at (3, 1.5) {\tt bac};
\node[vertex] (d) at (1.5, 0.5) {\tt ba};
\draw[->,dashed] (a) -- (b);
\draw[->,dashed] (b) -- (c);
\draw[->] (a) -- (d);
\draw[->] (d) -- (c);
\end{scope}
\end{scope}
\end{mypic}
\caption{Collapsing a~pair of arcs is replacing a~pair of dashed arcs with a~pair of solid arcs: general case (left) and example (right). The~``physical meaning'' of this transformation is that to get {\tt bac} from {\tt aba} one needs to cut~{\tt a} from the beginning and append~{\tt c} to the end and these two operations commute.}
\label{fig:collapsing}
\end{figure}

In this section, we describe a~natural way of normalizing an~Eulerian solution~$D$. Informally, it can be viewed as follows.
Imagine that all arcs of~$D$ form one~circular thread, and that there is a~nail in every node~$s \in {\cal S}$ corresponding to an input string. We apply {\em ``gravitation''} to the thread, i.e., we replace every pair of arcs $(\pref(v), v)$, $(v, \suff(v))$ 
with a~pair $(\pref(v), \pref(\suff(v)))$, $(\pref(\suff(v)), \suff(v))$, if there is no nail in~$v$ and if this does not disconnect~$D$. We call this {\em collapsing}, see Figure~\ref{fig:collapsing}.

A~formal pseudocode of the collapsing procedure is given in Algorithm~\ref{alg:collapse}. The pseudocode, in particular, reveals an important exception (not covered in Figure~\ref{fig:collapsing}):
if $|v|=1$, then $\pref(\suff(v))$ is undefined and we just remove the pair of arcs $(\pref(v), v)$ and $(v, \suff(v))$.

\begin{algorithm}[!ht]
\caption{Collapse}\label{alg:collapse}
\hspace*{\algorithmicindent} \textbf{Input:} hierarchical graph $HG(V,E)$, Eulerian solution~$D$, node~$v \in V$.
\begin{algorithmic}[1]
\If{$(\pref(v), v), (v, \suff(v)) \in D$}
\State\label{alg:col} $D \gets D \setminus \{(\pref(v), v), (v, \suff(v))\}$
\If{$|v| > 1$}
\State $D \gets D \cup \{(\pref(v), \pref(\suff(v))), (\pref(\suff(v)), \suff(v))\}$
\EndIf
\EndIf
\end{algorithmic}
\end{algorithm}

Algorithm~\ref{alg:ca}, that we call Collapsing Algorithm (CA), uses the property described above to normalize any solution.
It drops down all pairs of arcs that are not needed for connectivity.
\ab{(Recall that a~set of edges is called an Eulerian solution if it is connected and goes through all initial nodes and $\varepsilon$.)}


\begin{algorithm}[!ht]
\caption{Collapsing Algorithm (CA)}\label{alg:ca}
\hspace*{\algorithmicindent} \textbf{Input:} set of strings~$\mathcal{S}$, Eulerian solution~$D$ in~$HG$.\\
\hspace*{\algorithmicindent} \textbf{Output:} Eulerian solution $D'\colon |D'|\leq|D|$
\begin{algorithmic}[1]
\For{level~$l$ in $HG$ in descending order}\label{alg:ca_for}
\For{all $v \in V$ s.t. $|v|=l$ in lexicographic order:}
\While{$(\pref(v), v), (v, \suff(v)) \in D$ and collapsing it keeps~$D$ an Eulerian solution}
\State $\text{\sc Collapse}(HG, D, v)$
\EndWhile
\EndFor
\EndFor
\State return $D$
\end{algorithmic}
\end{algorithm}

\begin{figure}[ht]
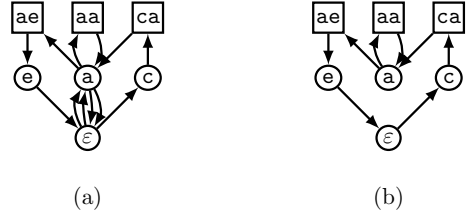

\begin{mypic}
\foreach \n/\x/\y in {ae/1/2, aa/2/2, ca/3/2}
  \node[inputvertex] (\n) at (\x,\y) {\tt \n};

\foreach \n/\x/\y in {a/2/1, e/1/1, c/3/1}
  \node[vertex] (\n) at (\x,\y) {\tt \n};

\node[vertex] (eps) at (2,0) {$\varepsilon$};
\node at (2,-1) {(a)};

\path (aa) edge[->,bend left] (a);
\path (a)    edge[->,bend left] (aa);
\path (eps) edge[->,bend left] (a);
\path (a)    edge[->,bend left] (eps);
\path (eps) edge[->,bend left=10] (a);
\path (a)    edge[->,bend left=10] (eps);

\foreach \f/\t in {a/ae, ae/e, e/eps, eps/c, c/ca, ca/a}
  \draw[->] (\f) -- (\t);
  
\begin{scope}[xshift=50mm]
\foreach \n/\x/\y in {ae/1/2, aa/2/2, ca/3/2}
  \node[inputvertex] (\n) at (\x,\y) {\tt \n};

\foreach \n/\x/\y in {a/2/1, e/1/1, c/3/1}
  \node[vertex] (\n) at (\x,\y) {\tt \n};

\node[vertex] (eps) at (2,0) {$\varepsilon$};
\node at (2,-1) {(b)};

\path (aa) edge[->,bend left] (a);
\path (a)    edge[->,bend left] (aa);

\foreach \f/\t in {a/ae, ae/e, e/eps, eps/c, c/ca, ca/a}
  \draw[->] (\f) -- (\t);
\end{scope}
\end{mypic}
\caption{(a)~An~Eulerian solution corresponding to the~permutation $({\tt ae}, {\tt aa}, {\tt ca})$. (b)~The solution from~(a) after normalization results in a shorter solution corresponding to the~permutation~$({\tt ca}, {\tt aa}, {\tt ae})$. \ab{This example also shows that although collapsing a~pair of edges is a~local change in the graph, it may drastically change the resulting superstring. In this case, it replaces a~superstring {\tt aeaaca} with a~shorter superstring {\tt caae}.}}
\label{fig:abnormalzigzag}
\end{figure}

It is easy to show (we prove this formally in Claim~\ref{claim:zigzag} on page~\pageref{claim:zigzag}) that any normalized solution is of the form~\eqref{eq:zigzag}. But it is not true that every zig-zag solution of the form~\eqref{eq:zigzag} is a normalized solution: see Figure~\ref{fig:abnormalzigzag} for an example. The normalization procedure does not just turn a~solution into some standard form, but it may also decrease its length.

\section{Collapsing Conjecture}
\label{sec:collapsing}

We are now ready to conjecture an astonishing structural property of the hierarchical graph: 
\begin{quote}
Take any Eulerian solution, double every arc of it, and normalize the resulting solution; the result is the same for all initial solutions!
\end{quote}
For the formal statement of the conjecture we use the following notation:  If $U$ and $V$ are two multisets, then $U\sqcup V$ is the multiset~$W$ such that each $w\in W$ has multiplicity equal to the sum of multiplicities it has in sets $U$ and $V$.
Formally, the conjecture is stated as follows.
\newtheorem*{scs}{Collapsing Conjecture}
\begin{scs}
For any set of strings ${\cal S}$ and any two Eulerian solutions~$D_1, D_2$ of~${\cal S}$, 
\begin{align*}
CA({\cal S}, D_1 \sqcup D_1) =  CA({\cal S}, D_2 \sqcup D_2) \, .
\end{align*}
\end{scs}

\begin{figure}[!ht]
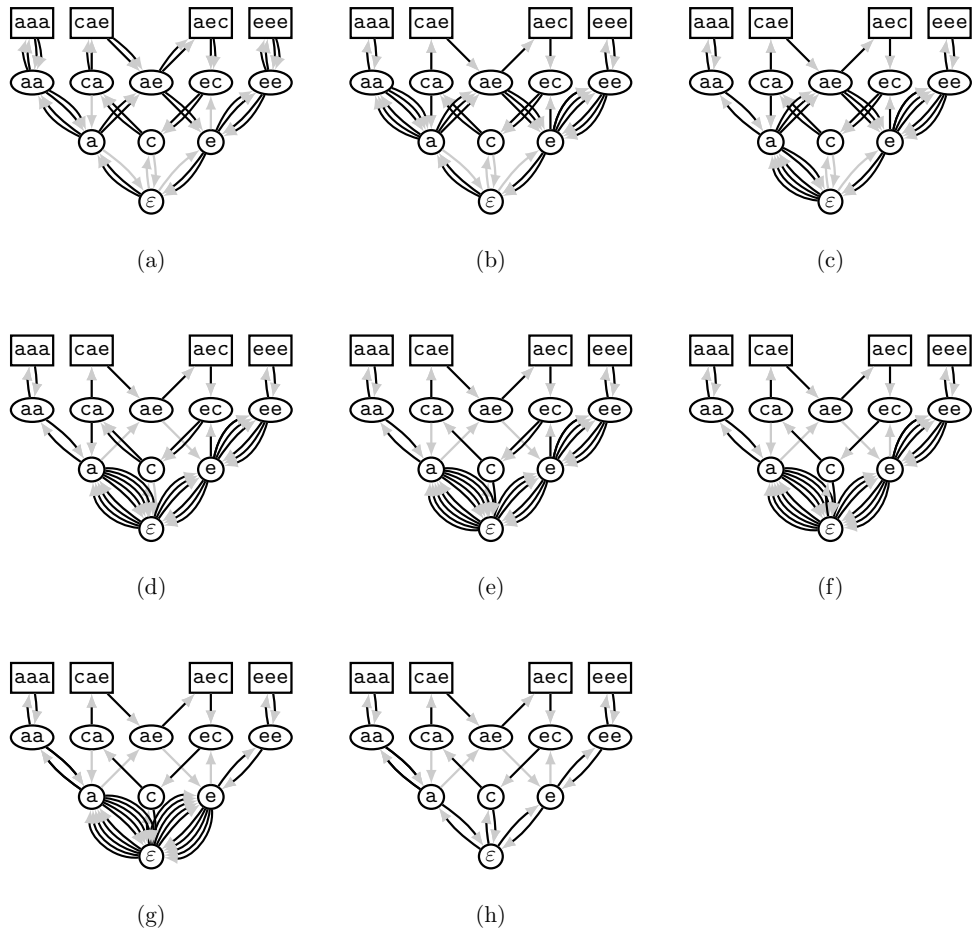

\begin{mypic}
\begin{scope}[scale=0.99,transform shape]
\we{0}{0}{a}{
\foreach \f/\t/\a in {eps/a/10, eps/a/20, 
a/aa/10, a/aa/20, 
aa/aaa/10, aa/aaa/20,
aaa/aa/10, aaa/aa/20,
aa/a/10, aa/a/20, 
a/ae/0, a/ae/10,
ae/aec/0, ae/aec/10,
aec/ec/0, aec/ec/10,
ec/c/0, ec/c/10,
c/ca/0, c/ca/10,
ca/cae/0, ca/cae/10,
cae/ae/0, cae/ae/10,
ae/e/0, ae/e/10,
e/ee/10, e/ee/20,
ee/eee/10, ee/eee/20,
eee/ee/10, eee/ee/20,
ee/e/10, ee/e/20,
e/eps/10, e/eps/20}
  \path (\f) edge[hgedge,bend left=\a,draw=black,thick] (\t);
}

\we{57}{0}{b}{
\foreach \f/\t/\a in {eps/a/10, eps/a/20, 
a/aa/10, a/aa/20, a/aa/30, aa/a/30,
aa/aaa/10,
aaa/aa/10,
aa/a/10, aa/a/20, 
a/ae/0, a/ae/10,
ae/aec/0, ae/e/20,
aec/ec/0, e/ec/0,
ec/c/0, ec/c/10,
c/ca/0, c/ca/10,
ca/cae/0, ca/a/0, a/ae/20,
cae/ae/0,
ae/e/0, ae/e/10,
e/ee/10, e/ee/20, e/ee/30,
ee/eee/10,
eee/ee/10,
ee/e/10, ee/e/20, ee/e/30,
e/ee/30,
e/eps/10, e/eps/20}
  \path (\f) edge[hgedge,bend left=\a,draw=black,thick] (\t);
}

\we{114}{0}{c}{
\foreach \f/\t/\a in {
eps/a/10, eps/a/20, eps/a/30, eps/a/40, a/eps/10, a/eps/20,
a/aa/10,
aa/aaa/10,
aaa/aa/10,
aa/a/10, 
a/ae/0, a/ae/10,
ae/aec/0, ae/e/20,
aec/ec/0, e/ec/0,
ec/c/0, ec/c/10,
c/ca/0, c/ca/10,
ca/cae/0, ca/a/0, a/ae/20,
cae/ae/0,
ae/e/0, ae/e/10,
e/ee/10, e/ee/20, e/ee/30,
ee/eee/10,
eee/ee/10,
ee/e/10, 
ee/e/20,
ee/e/30,
e/ee/30,
e/eps/10, e/eps/20}
  \path (\f) edge[hgedge,bend left=\a,draw=black,thick] (\t);
}

\we{0}{-55}{d}{
\foreach \f/\t/\a in {
eps/a/10, eps/a/20, eps/a/30, eps/a/40, 
a/eps/10, a/eps/20, a/eps/30, a/eps/40, a/eps/50,
a/aa/10,
aa/aaa/10,
aaa/aa/10,
aa/a/10, 
ae/aec/0,
aec/ec/0, e/ec/0,
ec/c/0, ec/c/10,
c/ca/0, c/ca/10,
ca/cae/0, ca/a/0,
cae/ae/0,
e/ee/10, e/ee/20, e/ee/30,
ee/eee/10,
eee/ee/10,
ee/e/10, ee/e/20, ee/e/30,
e/ee/30,
e/eps/10, e/eps/20, e/eps/30,
eps/e/10, eps/e/20, eps/e/30}
  \path (\f) edge[hgedge,bend left=\a,draw=black,thick] (\t);
}

\we{57}{-55}{e}{
\foreach \f/\t/\a in {
eps/a/10, eps/a/20, eps/a/30, eps/a/40, eps/a/50,
a/eps/10, a/eps/20, a/eps/30, a/eps/40, a/eps/50,
a/aa/10,
aa/aaa/10,
aaa/aa/10,
aa/a/10, 
ae/aec/0,
aec/ec/0, e/ec/0,
ec/c/0, ec/c/10,
c/ca/0,
ca/cae/0,
cae/ae/0,
e/ee/10, e/ee/20, e/ee/30,
ee/eee/10,
eee/ee/10,
ee/e/10, ee/e/20, ee/e/30,
e/ee/30,
e/eps/10, e/eps/20, e/eps/30,
eps/e/10, eps/e/20, eps/e/30,
c/eps/10}
  \path (\f) edge[hgedge,bend left=\a,draw=black,thick] (\t);
}

\we{114}{-55}{f}{
\foreach \f/\t/\a in {
eps/a/10, eps/a/20, eps/a/30, eps/a/40, eps/a/50,
a/eps/10, a/eps/20, a/eps/30, a/eps/40, a/eps/50,
a/aa/10,
aa/aaa/10,
aaa/aa/10,
aa/a/10, 
ae/aec/0,
aec/ec/0,
ec/c/0,
c/ca/0,
ca/cae/0,
cae/ae/0,
e/ee/10, e/ee/20, e/ee/30,
ee/eee/10,
eee/ee/10,
ee/e/10, ee/e/20, ee/e/30,
e/ee/30,
e/eps/10, e/eps/20, e/eps/30, e/eps/40,
eps/e/10, eps/e/20, eps/e/30,
c/eps/10, eps/c/10}
  \path (\f) edge[hgedge,bend left=\a,draw=black,thick] (\t);
}

\we{0}{-110}{g}{
\foreach \f/\t/\a in {aa/aaa/10, aaa/aa/10, ca/cae/0, cae/ae/0, ae/aec/0, aec/ec/0, ee/eee/10, eee/ee/10, aa/a/10, a/aa/10, c/ca/0, ec/c/0, ee/e/10, e/ee/10, a/aa/10, aa/a/10, eps/c/10, c/eps/10, 
e/eps/10, eps/e/10, 
e/eps/20, eps/e/20,
e/eps/30, eps/e/30,
e/eps/40, eps/e/40,
e/eps/50, eps/e/50,
a/eps/10, eps/a/10,
a/eps/20, eps/a/20,
a/eps/30, eps/a/30,
a/eps/40, eps/a/40,
a/eps/50, eps/a/50}
\path (\f) edge[hgedge,bend left=\a,draw=black,thick] (\t);
}

\we{57}{-110}{h}{
\foreach \f/\t/\a in {aa/aaa/10, aaa/aa/10, ca/cae/0, cae/ae/0, ae/aec/0, aec/ec/0, ee/eee/10, eee/ee/10, aa/a/10, a/aa/10, c/ca/0, ec/c/0, ee/e/10, e/ee/10, a/aa/10, aa/a/10, eps/c/10, c/eps/10, 
e/eps/10, eps/e/10, 
a/eps/10, eps/a/10}
\path (\f) edge[hgedge,bend left=\a,draw=black,thick] (\t);
}
\end{scope}
\end{mypic}
\caption{Stages of applying the Collapsing Algorithm to the dataset $\{{\tt aaa}, {\tt cae}, {\tt aec}, {\tt eee}\}$ and its \textbf{optimal} solution. (a)~We start by doubling every arc of the optimal solution from Figure~\ref{fig:hgex}(c). 
(b)~After collapsing all nodes at level $l=3$. 
(c)~After processing the node {\tt aa} at level $l=2$. Note that the algorithm leaves a~pair of arcs $({\tt a}, {\tt aa}), ({\tt aa}, {\tt a})$ as they are needed to connect the component $\{{\tt aa}, {\tt aaa}\}$ to the rest of the solution. (d)~After processing the {\tt ae} node. The algorithm collapses all pairs of arcs for this node as it lies in the same component as the~node~{\tt c}. 
(e)~After processing the {\tt ca} node.
(f)~After processing the {\tt ec} node.
(g)~After processing the {\tt ee} node. Note that at this point the solution has exactly the same length as at the very beginning (at stage~(a)).
(h)~Finally, after collapsing all the unnecessary pairs of arcs from the level~$l=1$. 
}
\label{fig:coll}
\end{figure}

\begin{figure}[!ht]
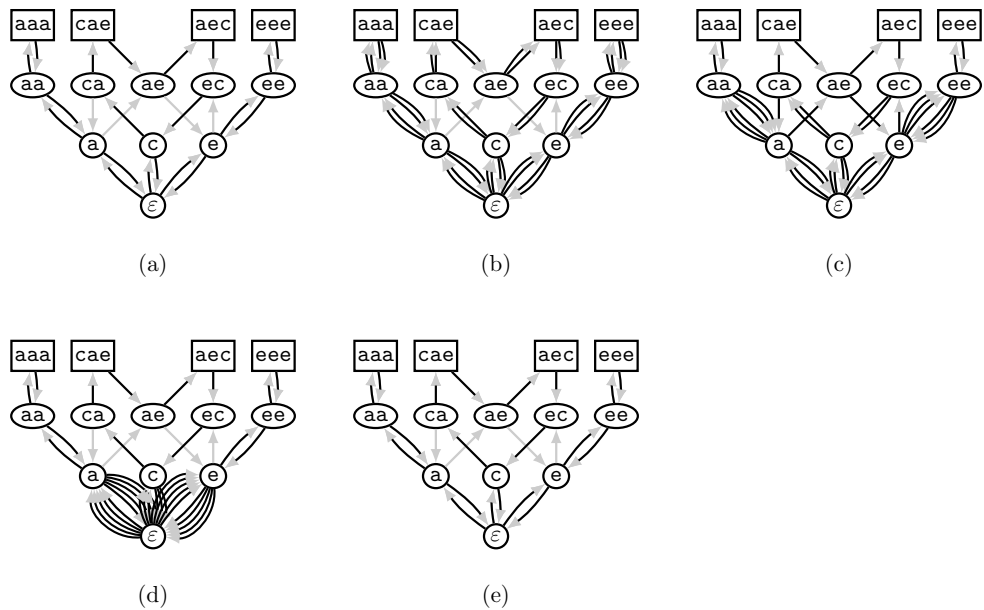

\begin{mypic}
\we{0}{0}{a}{
\foreach \f/\t/\a in {
eps/a/10,
a/aa/10,
aa/aaa/10,
aaa/aa/10,
aa/a/10,
a/eps/10,
eps/c/10,
c/ca/0,
ca/cae/0,
cae/ae/0,
ae/aec/0,
aec/ec/0,
ec/c/0,
c/eps/10,
eps/e/10,
e/ee/10,
ee/eee/10,
eee/ee/10,
ee/e/10,
e/eps/10
}
  \path (\f) edge[hgedge,bend left=\a,draw=black,thick] (\t);
}

\we{57}{0}{b}{
\foreach \f/\t/\a in {
eps/a/10, eps/a/20,
a/aa/10, a/aa/20,
aa/aaa/10, aa/aaa/20,
aaa/aa/10, aaa/aa/20,
aa/a/10, aa/a/20,
a/eps/10, a/eps/20,
eps/c/10, eps/c/20,
c/ca/0, c/ca/10,
ca/cae/0, ca/cae/10,
cae/ae/0, cae/ae/10,
ae/aec/0, ae/aec/10,
aec/ec/0, aec/ec/10,
ec/c/0, ec/c/10,
c/eps/10, c/eps/20,
eps/e/10, eps/e/20,
e/ee/10, e/ee/20,
ee/eee/10, ee/eee/20,
eee/ee/10, eee/ee/20,
ee/e/10, ee/e/20,
e/eps/10, e/eps/20
}
  \path (\f) edge[hgedge,bend left=\a,draw=black,thick] (\t);
}

\we{114}{0}{c}{
\foreach \f/\t/\a in {
eps/a/10, eps/a/20,
a/aa/10, a/aa/20, a/aa/30,
aa/aaa/10,
aaa/aa/10,
aa/a/10, aa/a/20, aa/a/30,
a/eps/10, a/eps/20,
eps/c/10, eps/c/20,
c/ca/0, c/ca/10,
ca/cae/0, ca/a/0,
cae/ae/0, a/ae/0,
ae/aec/0, ae/e/0,
aec/ec/0, e/ec/0,
ec/c/0, ec/c/10,
c/eps/10, c/eps/20,
eps/e/10, eps/e/20,
e/ee/10, e/ee/20,
ee/eee/10, ee/e/30,
eee/ee/10, e/ee/30,
ee/e/10, ee/e/20,
e/eps/10, e/eps/20
}
  \path (\f) edge[hgedge,bend left=\a,draw=black,thick] (\t);
}

\we{0}{-55}{d}{
\foreach \f/\t/\a in {
eps/a/10, eps/a/20, eps/a/30, eps/a/40, eps/a/50,
a/aa/10,
aa/aaa/10,
aaa/aa/10,
aa/a/10,
a/eps/10, a/eps/20, a/eps/30, a/eps/40, a/eps/50,
eps/c/10, eps/c/20, eps/c/30,
c/ca/0,
ca/cae/0,
cae/ae/0,
ae/aec/0,
aec/ec/0,
ec/c/0,
c/eps/10, c/eps/20, c/eps/30,
eps/e/10, eps/e/20, eps/e/30, eps/e/40, eps/e/50,
e/ee/10, 
ee/eee/10, 
eee/ee/10, 
ee/e/10, 
e/eps/10, e/eps/20, e/eps/30, e/eps/40, e/eps/50
}
  \path (\f) edge[hgedge,bend left=\a,draw=black,thick] (\t);
}

\we{57}{-55}{e}{
\foreach \f/\t/\a in {
eps/a/10, 
a/aa/10,
aa/aaa/10,
aaa/aa/10,
aa/a/10,
a/eps/10, 
eps/c/10, 
c/ca/0,
ca/cae/0,
cae/ae/0,
ae/aec/0,
aec/ec/0,
ec/c/0,
c/eps/10, 
eps/e/10, 
e/ee/10, 
ee/eee/10, 
eee/ee/10, 
ee/e/10, 
e/eps/10
}
  \path (\f) edge[hgedge,bend left=\a,draw=black,thick] (\t);
}

\end{mypic}
\caption{Stages of applying the Collapsing Algorithm to the dataset $\{{\tt aaa}, {\tt cae}, {\tt aec}, {\tt eee}\}$ and its \textbf{naive} solution resulting from overlapping the input strings in the same order as they are given. (a)~The solution of length 10 corresponding to the superstring {\tt aaacaeceee}. (b)~The doubled solution. (c)~After collapsing the $l=3$ level. (d)~After collapsing the $l=2$ level. (e)~After collapsing the $l=1$ level. 
}
\label{fig:collnaive}
\end{figure}

Figures~\ref{fig:coll} and~\ref{fig:collnaive} illustrate the action of
the Collapsing Algorithm for optimal and naive solutions, respectively. Note that the resulting solutions are equal. When processing level $l>1$ nodes, 
the collapsing procedure does not change the total length of the solution. What one normally sees at the beginning of the
$l=1$ iteration is an~Eulerian solution with many 
redundant pairs of arcs of the form $({\tt a}, \varepsilon)$, $(\varepsilon, {\tt a})$. It is exactly this stage of the algorithm where the total length of a~solution is decreased by the Collapsing Algorithm. 

We have verified the conjecture on millions of datasets (both handcrafted and randomly generated), and we invite the reader to see its visualizations and to check the conjecture on arbitrary datasets at the webpage~\cite{webpage}. 
Moreover, 
we support the conjecture by proving that it holds in the (NP-hard) special case where the input strings have length at most~3 in Section~\ref{subsec:scs3}.

If the Collapsing Conjecture is true, then there is a~simple and natural 2-approximate algorithm for SCS: take {\em any} Eulerian solution (e.g., merge the input strings in arbitrary order), double it, and apply the Collapsing Algorithm. Under the conjecture, this results in the same Eulerian solution as for doubled optimal solution and hence the length of the result is at most twice the optimal length.

\section{Greedy Hierarchical Conjecture}
\label{sec:greedy_hier}
\begin{figure}[!ht]
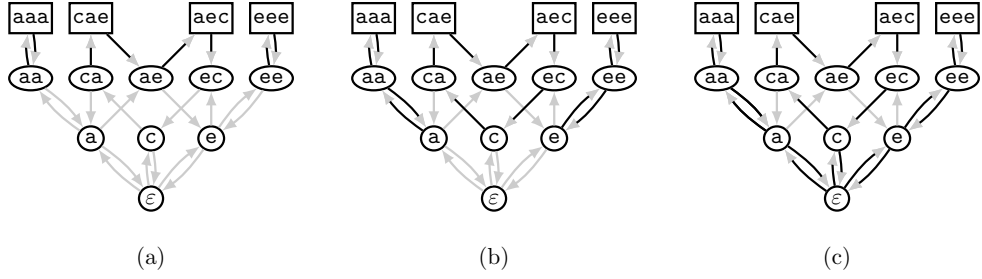

\begin{mypic}
\we{0}{0}{a}{
\foreach \f/\t/\a in {aa/aaa/10, aaa/aa/10, ca/cae/0, cae/ae/0, ae/aec/0, aec/ec/0, ee/eee/10, eee/ee/10}
  \path (\f) edge[hgedge,bend left=\a,draw=black,thick] (\t);
}

\we{57}{0}{b}{
\foreach \f/\t/\a in {aa/aaa/10, aaa/aa/10, ca/cae/0, cae/ae/0, ae/aec/0, aec/ec/0, ee/eee/10, eee/ee/10, aa/a/10, a/aa/10, c/ca/0, ec/c/0, ee/e/10, e/ee/10}
  \path (\f) edge[hgedge,bend left=\a,draw=black,thick] (\t);
}

\we{114}{0}{c}{
\foreach \f/\t/\a in {aa/aaa/10, aaa/aa/10, ca/cae/0, cae/ae/0, ae/aec/0, aec/ec/0, ee/eee/10, eee/ee/10, aa/a/10, a/aa/10, c/ca/0, ec/c/0, ee/e/10, e/ee/10, a/aa/10, aa/a/10, eps/c/10, c/eps/10, e/eps/10, eps/e/10, a/eps/10, eps/a/10}
\path (\f) edge[hgedge,bend left=\a,draw=black,thick] (\t);
}
\end{mypic}
\caption{(a)~After processing the $l=3$ level. (b)~After processing the $l=2$ level. Note that for the node {\tt aa} we add two lower arcs ($({\tt a}, {\tt aa})$ and $({\tt aa}, {\tt a})$) since otherwise the corresponding weakly connected component ($\{{\tt aa}, {\tt aaa}\}$) will not be connected to the rest of the solution. At the same time, when processing the node {\tt ae} we observe that it lies in a~weakly connected component that contains imbalanced nodes ({\tt ca} and {\tt ec}), hence there is no need to add two lower arcs to {\tt ae}. (c)~After processing the $l=1$ level. The resulting solution has length~10 and is, therefore, suboptimal (compare it with the optimal solution shown in Figure~\ref{fig:hgex}(c)).}
\label{fig:hgexa}
\end{figure}

In this section, we present one more curious property of the Collapsing Algorithm that reveals its intricate connection to greedy algorithms. For this, we introduce the so called Greedy Hierarchical Algorithm (GHA) that constructs an Eulerian solution in a~stingy fashion, i.e., tries to add as few arcs as possible:
\begin{quote}
Proceed nodes from top to bottom. For each node, ensure that it is balanced and connected to the rest of the solution.
\end{quote}
This is best illustrated with an example, see Figure~\ref{fig:hgexa}. We start constructing an~Eulerian solution~$D$ by processing the nodes at level~$3$. The solution~$D$ must visit all these four nodes, so we add all incoming and outgoing arcs to~$D$, see Figure~\ref{fig:hgexa}(a). We then process the level~2. The node~{\tt aa} is balanced, but if we skip it, it will not be connected to the rest of the solution, so we add to~$D$ the arcs $({\tt a}, {\tt aa})$ and $({\tt aa}, {\tt a})$. The node {\tt ae} is balanced, we do nothing for it. The node {\tt ca} is imbalanced, so we add an arc $({\tt c}, {\tt ca})$ to~$D$. We balance the node {\tt ec} similarly. The node {\tt ee} is processed similarly to the node {\tt aa}. The result of processing the second level is shown in Figure~\ref{fig:hgexa}(b). On the last stage we connect the nodes {\tt a}, {\tt b}, and {\tt c} to $\varepsilon$ to ensure connectivity, see Figure~\ref{fig:hgexa}(c). Hence, when processing level~$l$, we only add arcs between levels~$l$ and~$l-1$.

More formally, GHA first considers the input
strings~${\cal S}$. Since we assume that 
no $s \in {\cal S}$ is a~substring of another 
$t \in {\cal S}$, there is no down-path from~$t$ to~$s$ in $HG$. 
This means that any walk through $\varepsilon$ and ${\cal S}$ goes through the arcs $\{(\operatorname{pref}(s), s), (s, \operatorname{suff}(s)) \colon s \in {\cal S}\}$. The algorithm adds all of them to the constructed Eulerian solution~$D$ and starts processing all the nodes level by level, from top to bottom. At each level, we process the nodes in the lexicographic order. If the degree of the current node~$v$ is imbalanced, we balance it by adding an appropriate number of incoming (i.e., $(\pref(v),v)$) or outgoing (i.e., $(v, \suff(v))$) arcs from the previous (i.e., lower) level. In the case when $v$~is balanced, we just skip it. The only exception when we cannot skip it is when {\em $v$~lies in an Eulerian component and $v$ is the last chance of this component to be connected to the rest of the arcs in~$D$}. (See, for example, the vertex $ {\tt aa}$ in Figure~\ref{fig:hgexa}(a)). The pseudocode is given in~Algorithm~\ref{algo:gha}. 

\begin{algorithm}[!ht]
\caption{Greedy Hierarchical Algorithm (GHA)}\label{algo:gha}
\hspace*{\algorithmicindent} \textbf{Input:} set of strings~${\cal S}$.\\
\hspace*{\algorithmicindent} \textbf{Output:} Eulerian solution~$D$.
\begin{algorithmic}[1]
\State $HG(V,E) \gets \text{hierarchical graph of ${\cal S}$}$ 
\State\label{alg:gha_init}$D \gets \{(\operatorname{pref}(s), s), (s, \operatorname{suff}(s)) \colon s \in {\cal S}\}$
\For{level $l$ from $\max\{|s| \colon s \in {\cal S}\}$ downto 1}\label{alg:for}
\For{node $v \in V$ with $|v|=l$ in the lexicographic order}
\If{$|\{(u,v) \in D \colon |u|=|v|+1\}| \neq |\{(v,w) \in D \colon |w|=|v|+1\}| $}
\State\label{alg:step6} balance the degree of $v$ in~$D$ by adding an appropriate number of lower arcs
\Else
\State\label{alg:else} ${\cal C} \gets \text{weakly connected component of $v$ in $D$}$
\State $u \gets \text{the lexicographically largest string among shortest strings in ${\cal C}$}$
\If{${\cal C}$ is Eulerian, $\varepsilon \not \in {\cal C}$, and $v = u$}
\State\label{alg:last} $D \gets D \cup \{(\pref(v), v), (v, \suff(v))\}$
\EndIf
\EndIf
\EndFor
\EndFor
\State return $D$
\end{algorithmic}
\end{algorithm}

While GHA is almost as simple as the standard Greedy Algorithm (GA), GHA has several provable advantages over GA:
\begin{description}
\item One advantage of GHA over GA is that GHA is more flexible in the following sense. On every step, GA selects two strings and fixes tightly their order. GHA instead works to ensure connectivity. When the resulting set~$D$ is connected, an actual order of input strings is given by the corresponding Eulerian cycle through~$D$. This is best illustrated on the following~toy example. For the~dataset $\mathcal{S}=\{{\tt ae}, {\tt ea}, {\tt ee}\}$, GA might produce a~suboptimal solution {\tt aeaee} if it merges the strings {\tt ae} and {\tt ea} at the first step. At the same time, it is not difficult to see that GHA finds an optimal solution for~$\mathcal{S}$.
\item Another advantage of GHA is that, in contrast to GA, it solves {\em exactly} two well known polynomially solvable special cases of SCS: when the input strings have length at most two and when the input strings form a~$k$-spectrum of an unknown string (that is, the input strings constitute all
$k$-substrings of a~string). We prove this formally in Sections~\ref{sec:ghatwo} and~\ref{sec:ghaspectrum}. Informally, this happens because for such datasets there are no connectivity issues for GHA: for $k$-spectrum, after processing the highest level GHA gets a~weakly connected component; for 2-SCS, after processing the level~2, GHA gets several weekly connected components such that different components do not share common letters and therefore are completely independent. Figure~\ref{fig:abnormalzigzag}(b) illustrates this: while GA may produce a~permutation $({\tt ca}, {\tt ae}, {\tt aa})$, GHA constructs an optimal
permutation $({\tt ca}, {\tt aa}, {\tt ae})$.

In Section~\ref{sec:ghatough}, we also show a~dataset where GHA produces a~solution that is almost two times longer than the optimal one.

\item In Section~\ref{sec:gr_im_wghc}, \ab{we show that the approximation guarantee of GHA is no worse than that of GA. 
Combining with the result of Kaplan and Shafrir~\cite{KS2005},
this implies immediately that GHA is 3.5-approximate.}
Moreover, we prove that the standard Greedy Conjecture implies $2$-approximation of GHA, which makes it natural to study the approximation ratio of GHA.
\end{description}

We are now ready to state our second conjecture: the results of the Collapsing Algorithm and Greedy Hierarchical Algorithm coincide!
\newtheorem*{ghcc}{Greedy Hierarchical Conjecture}
\begin{ghcc}
For any set of strings~$\mathcal{S}$ and any Eulerian solution~$D$,
\[CA(\mathcal{S}, D \sqcup D) = GHA(\mathcal{S}) \, .\]
\end{ghcc}
While the Greedy Hierarchical Conjecture implies that GHA finds a $2$-approximate solution, we separately state this weak version of the conjecture.

\newtheorem*{wghcc}{Weak Greedy Hierarchical Conjecture}
\begin{wghcc}
GHA is a factor $2$ approximation algorithm for the Shortest Common Superstring problem.
\end{wghcc}
\section{Relations between the Conjectures}
\begin{figure}
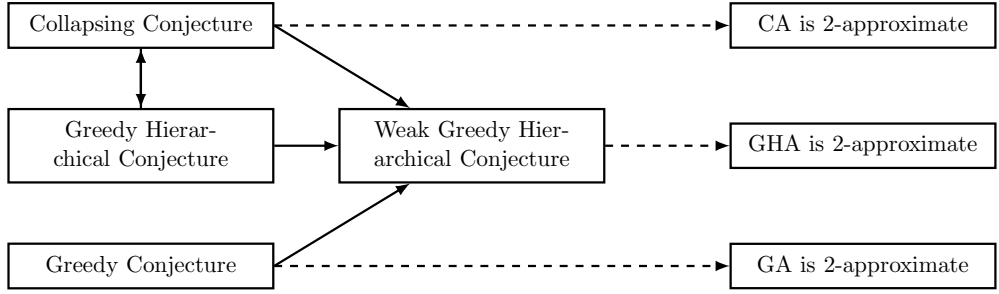

\begin{mypic}
\tikzstyle{r}=[rectangle,inner sep=2mm,draw,text centered, text width=40mm]

\node[r] (w) at (5.5,3) {Weak Greedy Hierarchical Conjecture};
\node[r] (cc) at (0,5) {Collapsing Conjecture}; 
\node[r] (gc) at (0,1) {Greedy Conjecture};
\node[r] (ghc) at (0,3) {Greedy Hierarchical Conjecture}; 

\node[r] (ca) at (12,5) {CA is 2-approximate};
\node[r] (gha) at (12,3) {GHA is 2-approximate};
\node[r] (ga) at (12,1) {GA is 2-approximate};


\foreach \f/\t in {cc/ghc, cc.east/w, ghc/w, gc.east/w, ghc/cc}
  \draw[->] (\f) -- (\t);
  
\foreach \f/\t in {cc/ca, w/gha, gc/ga}
  \draw[dashed,->] (\f) -- (\t);

\end{mypic}
\caption{Relations between the conjectures (left), and the 2-approximate algorithms they imply (right). 
Collapsing and Greedy Hierarchical Conjectures are equivalent. They imply the weak version of the Greedy Hierarchical Conjecture, which also follows from the standard Greedy Conjecture. Each conjecture implies that the corresponding algorithm finds a $2$-approximate solution for SCS.}
\label{fig:relations}
\end{figure}

In this section we prove some relations between the Collapsing and Greedy conjectures. Namely, in Section~\ref{sec:equiv} we prove the equivalence of Collapsing and Greedy Hierarchical conjectures. In Section~\ref{sec:gr_im_wghc} we prove that the standard Greedy Conjecture implies Weak Hierarchical Greedy Conjecutre (which is sufficient for a simple 2-approximate greedy algorithm for SCS). Finally, it is easy to see that Greedy Hierarchical Conjecture implies its weak version: indeed, if every doubled solution results in the solution obtained by GHA, then GHA does not exceed twice the optimal superstring length. In Figure~\ref{fig:relations}, we show the proven relations between the conjectures, together with 2-approximate algorithm which follow from each of the conjecture.
%
%
%
%

\subsection{Equivalence of Collapsing and Greedy Hierarchical Conjectures}
\label{sec:equiv}
In this section we prove the equivalence of Collapsing Conjecture and Greedy Hierarchical Conjecture. Recall that Collapsing Conjecture claims that for any pair of Eulerian solutions $D_1$ and $D_2$ for the input strings ${\cal S}$, we have 
\[CA({\cal S}, D_1 \sqcup D_1) =  CA({\cal S}, D_2 \sqcup D_2)\, .\]
 The Greedy Hierarchical Solution extends this statement to 
 \[CA({\cal S}, D_1 \sqcup D_1) =  GHA(\mathcal{S}) \, .\] Greedy Hierarchical Conjecture trivially implies Collapsing conjecture, and in order to prove their equivalence, it suffices to show that the collapsing procedure applied to the doubled GHA solution results in the GHA solution: 
 \[CA({\cal S}, GHA(\mathcal{S}) \sqcup GHA(\mathcal{S})) =  GHA(\mathcal{S}) \, .\]
 
 \begin{theorem}
For any set of strings $\mathcal{S}$,
\[
			CA(\mathcal{S}, GHA(\mathcal{S}) \sqcup GHA(\mathcal{S})) = GHA(\mathcal{S}) \, .
\]
\end{theorem}
\begin{proof}
Let us denote two copies of the $GHA(\mathcal{S})$ solution by $B$ and $R$, which stand for a blue-copy and a red-copy. We will prove the theorem statement by showing that $CA(R \sqcup B)$ collapses all arcs of $B$ and keeps untouched the arcs of $R$, as this implies that $CA(R \sqcup B)=R=GHA(\mathcal{S})$. For this, without loss of generality assume that the Collapsing Algorithm collapses blue arcs first, that is, if for a vertex $v$, $CA$ can collapse a blue pair of arcs $(\pref(v), v), (v, \suff(v))$, it does so. Recall that $CA$ processes vertices in the descending order of levels (Algorithm~\ref{alg:ca}, line~\ref{alg:ca_for}). We will prove that before processing level $l$, all the arcs above it (i.e., the arcs with at least one vertex at a level  $>l$) do satisfy the desired property: all blue arcs are collapsed, and all red arcs are untouched.

The base case trivially holds for $l := \max\{|s| \, : \, s\in\mathcal{S}\}$, since the set of arcs above the level $l$ is empty. Assume the claim is true for the level $k > 0$, and let us prove the claim for the level $k - 1$. Note that regardless of the number of collapse operations applied to $B$, $B$ remains a set of walks: indeed, the collapse procedure keeps the balance of incoming and outgoing arcs for each vertex. By the induction hypothesis all the blue arcs above the level $k$ are collapsed, so we have that if for a vertex $v$ at level $k$ there is an arc $(\pref(v), v)$ in $B$, then there is also an arc $(v, \suff(v))$ in $B$, and vice versa. Recall that $CA$ collapses blue arcs when possible, and since every vertex has the same number of blue incoming and outgoing arcs, all pairs collapsed at the level $k$ are monotone. 
	
Now let us show that no red pair can be collapsed. Indeed, if for some vertex $v$ at level $k$ there is a red pair $(\pref(v), v)$, $(v, \suff(v))$, then by construction of $GHA$ $v$ is either in $\mathcal{S}$ or is the last chance of the corresponding component $\mathcal{C} \ni v$ to be connected to the remaining arcs in $R$ (note that the first case is a subcase of the second one, as then $\mathcal{C}$ contains only one vertex). It follows that if $CA$ collapses such a pair of arcs, then $v$ has no blue arcs (as they have been collapsed before the red arcs), and all other vertices in the component $\mathcal{C}$ at the level $k$ collapsed all arcs (since $v$ is the last vertex in $\mathcal{C}$ in lexicographic order). Therefore, this pair is also the last chance of $\mathcal{C}$ to be connected to the rest of the arcs in $R$, thus, $CA$ cannot collapse it.
	
	It remains to show that all blue pairs at level $k$ are collapsed. This trivially holds because no red pair is collapsed, and, thus, the connectivity of $R \sqcup B$ is maintained by $R$. This finishes the proof.
\end{proof}

\subsection{Greedy Implies Greedy Hierarchical}
\label{sec:gr_im_wghc}
Consider a~permutation of the input strings. We say that it is a~{\em valid greedy permutation} if it can be constructed by the Greedy Algorithm: there exist $n-1$ merges of the $n$~input strings that lead to this permutation such that at every step the two merged strings have the largest overlap. We will prove that GHA always returns a solution which corresponds to a greedy permutation of the input string. That is, while the standard Greedy Algorithm does not determine how to break ties, the Greedy Hierarchical Algorithm is a specific instantiation of the Greedy Algorithm with some tie-breaking rule.

We will use the following simple property of solutions constructed by the GHA algorithm.
\begin{claim}
\label{claim:zigzag}
Let $D$ be an Eulerian solution constrtucted by GHA. Then $D$ has a ``zig-zag'' form as in~\eqref{eq:zigzag}. 
\end{claim}
\begin{proof}
First we prove that $D$ is normalized, that is, any application of the collapsing procedure of Algorithm~\ref{alg:collapse} to $D$ will violate the property of Eulerian solution. Indeed, Algorithm~\ref{alg:collapse} can only collapse pairs of arcs of the form $(\pref(s), s), (s, \suff(s))$. The Greedy Hierarchical Algorithm adds such pairs to its solution in two cases: (i) $s$ is an input string (line~\ref{alg:gha_init} of Algorithm~\ref{algo:gha}); (ii) $s$ is the the lexicographically largest among the shortest strings in its Eulerian component (line~\ref{alg:last} of Algorithm~\ref{algo:gha}). Now note that in the former case, the collapsing procedure applied to $s$ would violate the property that $D$ must contain all input strings, and in the latter case, the collapsing procedure would violate the connectivity property of $D$.

We finish the proof by showing that every normalized solution is of the form~\eqref{eq:zigzag}. Let $\pi=(s_1, \dots, s_n)$ be  the permutation of the input strings corresponding to a normalized Eulerian solution~$D$. Let us follow the arcs of~$D$ in the order of the permutation $\pi$, and let $P$ be the set of arcs between the input strings $s_i$ and $s_{i+1}$. We will prove that $P$ is the union of the sets of arcs of the paths $s_i \to \overlap(s_i,s_{i+1})$ and $\overlap(s_i,s_{i+1})\to s_{i+1}$. If $P$ contains a pair of consecutive up- and down-arcs, that is, there exists a pair of arcs $(\pref(s), s), (s, \suff(s))$ in $P$, then this pair would have been collapsed by Algorithm~\ref{alg:collapse}, line~\ref{alg:col}. Therefore, the path $P$ consists of a number of down-arcs followed by a number of up-arcs. It remains to show that the number of down-arcs in $P$ is $d=|s_i|-|\overlap(s_i,s_{i+1})|$. Note that by the definition of $\overlap(\cdot,\cdot)$, the number of down-arcs in~$P$ is at least~$d$. On the other hand, if the number of down-arcs in~$P$ is strictly greater than~$d$, then both the down-path and up-path in~$P$ contain the vertex $\overlap(s_i, s_{i+1})$. This implies that the pair of arcs  $(\pref(s), s), (s, \suff(s))$ for $s=\overlap(s_i, s_{i+1})$ would have been collapsed by Algorithm~\ref{alg:collapse}, line~\ref{alg:col}, as it does not violate the connectivity of the solution. Therefore, the number of down-arcs in $P$ is exactly $d$, which implies that $P$ is the path $s_i \to \overlap(s_i,s_{i+1})$ followed by the path $\overlap(s_i,s_{i+1})\to s_{i+1}$.
\end{proof}

\begin{theorem}
\label{thm:gr_im_wghc}
Every permutation $\pi=(s_1, \dots, s_n)$ of the input strings constructed by GHA is a~valid greedy permutation.
\end{theorem}
\begin{proof}
Consider the following algorithm~$A$: it starts with the sequence $(s_1, \dots, s_n)$ obtained by GHA, and at every step it merges two neighboring strings in this sequence that have the largest overlap. It is a~greedy algorithm, but instead of considering all pairwise overlaps, it only considers overlaps of neighboring strings in the sequence. Of course, in the end, this algorithm constructs exactly the permutation~$\pi$. To show that $\pi$ is a~valid greedy permutation, we show that at every iteration of~$A$ no two strings have longer overlap than the two strings merged by~$A$.

Consider, for the sake of contradiction, the first iteration when the algorithm $A$~merges some pair of neighboring strings with overlap of length~$k$ whereas there are non-neighboring strings~$p$ and~$q$ with $v=\overlap(p,q)$, $|v|>k$. 
At this point, $p$~is a~merger of input strings $s_a, s_{a+1}, \dotsc, s_b$
and $q$~is a~merger of input strings $s_c, s_{c+1}, \dotsc, s_d$. 
Then, from the assumption that no input string contains another input string, we have that $v=\overlap(p,q)=\overlap(s_b,s_c)$. Since the algorithm~$A$
merges neighboring strings in the decreasing order of overlap lengths, we have that $|\overlap(s_b,s_{b+1})| \le k <|v|$ and $|\overlap(s_{c-1},c_c)| \le k < |v|$.\footnote{In the case when $s_b$ is the last string in the solution (or $s_c$ is the first string in the solution) we think of it being followed by $\varepsilon$, and $|\overlap(s_b,\varepsilon)|=0<|v|$ still holds.} 

Now we consider the Eulerian solution $D$ constructed by GHA in the hierarchical graph.
By Claim~\ref{claim:zigzag}, $D$ has a ``zig-zag'' form, thus, it contains all arcs from the path $s_b\to\overlap(s_{b}, s_{b+1}) \to s_{b+1}$, and all arcs from the path $s_{c-1}\to\overlap(s_{c-1}, s_{c}) \to s_{c}$. Recall that $v=\overlap(s_b,s_c)$, and that $|\overlap(s_b,s_{b+1})| <|v|$ and $|\overlap(s_{c-1},s_c)|< |v|$. In particular, the paths $s_b\to\overlap(s_{b}, s_{b+1})$ and $\overlap(s_{c-1}, s_{c}) \to s_{c}$ pass through the vertex $v$, which implies that the vertex $v$ in the solution $D$ has at least one incoming arc from the previous level and at least one outgoing arc to the previous level (see~Figure~\ref{fig:gagha}(a)). Such a~pair of arcs in the Eulerian solution $D$ constructed by GHA may only occur when~$v$ is the last chance of
its connected component to be connected to the rest of the solution (see line~\ref{alg:last} of Algorithm~\ref{algo:gha}). This, in turn, implies that right before the pair of arcs $(\pref(v), v)$ and $(v, \suff(v))$ was added to the Eulerian solution, there was an~Eulerian component where $v$~was the lexicographically largest among all shortest nodes. This component is shown schematically in~Figure~\ref{fig:gagha}(b). All overlap-nodes (the nodes which are equal to $\overlap(s_i,s_{i+1})$) of this component lie on levels~$\geq k$. Note that the pair $(\pref(v), v)$ and $(v, \suff(v))$ is added to the solution by GHA exactly once (line~\ref{alg:last} of Algorithm~\ref{algo:gha}). Therefore, any path following the arcs of $D$, after going through the arc $(\pref(v), v)$ must traverse the overlying component containing $s_b$ and $s_c$ (as otherwise the path could not reach the overlying component). In turn, this implies that after considering all overlaps of length $|v|>k$, $s_b$ and $s_c$ are already merged into one string, so they cannot be merged at this stage.

\begin{figure}
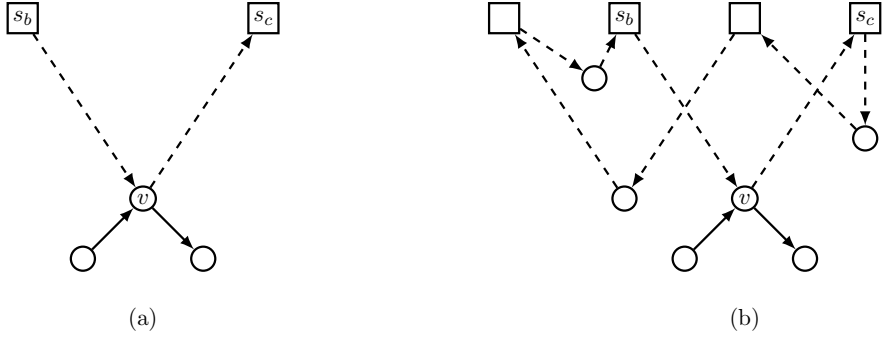

\begin{mypic}
\node[inputvertex] (b) at (0,6) {$s_b$}; 
\node[inputvertex] (c) at (4,6) {$s_c$}; 
\node[vertex] (v) at (2,3) {$v$};
\node[vertex] (pv) at (1,2) {};
\node[vertex] (sv) at (3,2) {};
\draw[->] (pv) -- (v);
\draw[->] (v) -- (sv);
\draw[->,anypath] (b) -- (v);
\draw[->,anypath] (v) -- (c);

\node at (2,1) {(a)};

\begin{scope}[xshift=100mm]
\node[inputvertex] (b) at (0,6) {$s_b$}; 
\node[inputvertex] (c) at (4,6) {$s_c$}; 
\node[inputvertex] (d) at (2,6) {}; 
\node[inputvertex] (e) at (-2,6) {}; 
\node[vertex] (f) at (4,4) {};
\node[vertex] (g) at (0,3) {};
\node[vertex] (h) at (-0.5,5) {};
\node[vertex] (v) at (2,3) {$v$};
\node[vertex] (pv) at (1,2) {};
\node[vertex] (sv) at (3,2) {};
\draw[->] (pv) -- (v);
\draw[->] (v) -- (sv);

\foreach \f/\t in {b/v, v/c, c/f, f/d, d/g, g/e, e/h, h/b}
  \draw[->,anypath] (\f) -- (\t); 

\node at (2,1) {(b)};
\end{scope}
\end{mypic}
\caption{(a)~In the Eulerian solution the node $v=\overlap(s_b,s_c)$ has a~pair of lower arcs. (b)~For this reason, above~$v$, there is an~Eulerian component.}
\label{fig:gagha}
\end{figure}
\end{proof}

Theorem~\ref{thm:gr_im_wghc} has two immediate corollaries.
\begin{corollary}
The Greedy Conjecture implies the Weak Greedy Hierarchical Conjecture: if the Greedy Algorithm is 2-approximate, then so is the Greedy Hierarchical Algorithm.
\end{corollary}
Since every valid greedy permutation is a $3.5$-approximation to the Shortest Common Superstring problem~\cite{KS2005}, we have the following corollary.
\begin{corollary}
GHA is a factor $3.5$ approximation algorithm for the Shortest Common Superstring problem.
\end{corollary}
\section{Proof of Collapsing Conjecture for Strings of Length~3}
\label{subsec:scs3}
In this section, we show that the Collapsing Conjecture holds for 
the special case
when input strings have length at most three. 
Remarkably, this follows from a~more general theorem stated below.

\begin{theorem}\label{thm:cc3}
Let $\mathcal{S}$ contain strings of length at most~3 and let $L$~be an Eulerian solution that for each $s \in \mathcal{S}$ contains at least two copies of arcs $(\pref(s), s)$ and $(s, \suff(s))$. Then $CA(\mathcal{S}, L)=GHA(\mathcal{S})$.
\end{theorem}

It is not difficult to see that the theorem indeed implies the Collapsing Conjecture: clearly, $L=D \sqcup D$, where $D$ is an Eulaerian solution, satisfies the condition. Moreover, this also works for $L=D_1 \sqcup D_2$, where $D_1, D_2$ are arbitrary Eulerian solutions, and for $L=D \sqcup CC$, where $CC$ is a~cycle cover, i.e., a~set of cycles that go through all input strings. The main difference between an Eulerian solution and a~cycle cover is that the later is not required to be connected. For this reason, any Eulerian solution is also a~cycle cover (but not vice versa) and hence an~{\em optimal} cycle cover is definitely not longer than an optimal Eulerian solution: $OPT \le OPTCC$. Hence, Theorem~\ref{thm:cc3} says that the result of GHA is not just no longer than $2\cdot OPT$, but even no longer than $OPT+OPTCC$.

Before proving Theorem~\ref{thm:cc3}, we introduce some notation and 
prove two auxiliary results. Recall that the Collapsing Algorithm processes the nodes level by level. Denote by $L_i$ an intermediate Eulerian solution right before is starts collapsing the nodes at level~$i$ (that is, in $L_i$ all the nodes at levels $>i$ are already collapsed). 
For an arbitrary Eulerian solution~$U$, by $\operatorname{above}(U,i)$ denote the part of~$U$ that lies above the level~$i$: 
$\operatorname{above}(U,i)=\{(u, v) \in U \colon |u|, |v| \ge i\}$.
We show that $\operatorname{above}(D,i)=\operatorname{above}(L_i, i)$ for every~$i$. This is enough since then \[CA(\mathcal{S}, L)=\operatorname{above}(L_0,0)=\operatorname{above}(D,0)=GHA(\mathcal{S}) \, .\]


\begin{lemma}\label{lem:path}
Let $w$~be a~walk from~$u$ to~$v$ in an Eulerian solution with all its nodes at levels~$\le k$. Consider a~single collapsing step for a~node~$t$ that is either an~{\em intermediate} node of~$p$ at level~$k$ or is a~node at level $<k$ that do not belong to~$p$. Then $w$~is still a~walk from~$u$ to~$v$ in the resulting solution. 
\end{lemma}
\begin{proof}
Indeed, if $t$~does not belong to~$w$, then collapsing it does not change~$w$ at all. Otherwise $t$~is an intermediate node of~$w$ at level~$k$. Since $w$~does not have any node above level~$k$, $w$~goes through
$(\pref(t), t), (t, \suff(t))$. Clearly, collapsing~$t$ keeps $w$~a~walk.
\end{proof}

\begin{lemma}\label{lemma:toepsilon}
Let $v$~be a~node in~$L_2$ at level~$1 \le l \le 2$ (i.e., $l=|v|$). Then there is a~walk from~$\varepsilon$ to~$v$ and a~walk from~$\varepsilon$ to~$v$ in~$L_l$ that does not contain nodes at level~3.
\end{lemma}
\begin{proof}
We start by proving that there is a~walk from~$v$ to~$\varepsilon$ 
for $|v|=2$ (the existence of a~walk from~$\varepsilon$ to~$v$ is proved in a~similar fashion).

Consider a~walk~$w$ from~$v$ to~$\varepsilon$ in~$L$ (there is such a~walk as $L$~is an~Eulerian solution). All repeated nodes in~$w$ may be removed, 
so one may assume that $w$~passes through its nodes at level~3 exactly once. Then, it is sufficient to show that each such node is collapsed.

Consider a~node~$s$ of~$w$ at level~3 and a~pair of arcs  $(\pref(s), s), (s, \suff(s)) \in w$ going through it. If $s$~is not at input string (i.e., $s \not \in \mathcal{S}$), then CA collapses this pair of arcs and this does not disconnect~$w$. On the other hand, if $s$~is an input string ($s \in \mathcal{S}$), then there are two copies of $(\pref(s), s), (s, \suff(s))$ in~$L$. At least one copy of this pair is collapsed in~$L$ and therefore belongs to~$L_2$.

The statement for~$v$ with $|v|=1$ follows from Lemma~\ref{lem:path}.
\end{proof}

\begin{proof}[Proof of Theorem~\ref{thm:cc3}]
As discussed above, it suffices to prove that $\operatorname{above}(D,i)=\operatorname{above}(L_i, i)$ for every $i=2,1,0$.

\begin{description}
\item {\em Level $i=2$.} The base case $i=2$ is straightforward: clearly, the Collapsing Algorithm leaves exactly one copy of arcs $(\pref(s), s)$ and $(s, \suff(s))$ for every $s \in \mathcal{S}$ and fully collapses all other nodes at level~$3$. Then, $\operatorname{above}(L_2,2)=\operatorname{above}(D,2)$ as $(\pref(s), s), (s, \suff(s))$ for $s \in \mathcal{S}$ are the only edges between levels~2 and~3 in $D$.

\item {\em Level $i=1$.} Note that $\operatorname{above}(L_2,2) \subseteq L_2$ and $L_2$ is an Eulerian cycle. Hence, $\operatorname{above}(L_2,2)$ is a~collection of walks. Consider such a~walk~$w$ and consider two cases.
\begin{itemize}
\item $w$ is a~closed walk. Let~$v$ be the lexicographically largest node of~$w$ at level~2. What we want to show is that in $L_1$ this closed walk~$w$ is connected to the rest of $L_1$ through a~pair of arcs $(\pref(v), v)$ $(v, \suff(v))$ only.

By Lemma~\ref{lemma:toepsilon}, there is a~path from $v$ to $\varepsilon$ in $L_2$ and hence $(v, \suff(v)) \in L_2$; similarly, $(\pref(v), v) \in L_2$. Since $v$~is lexicographically largest at level~$2$ in~$w$, when CA starts processing the node~$v$, all other nodes at level~2 in~$w$ are fully collapsed, i.e., for any such node~$u$, $(\pref(u), u) \not \in L_1$ and $(u, \suff(u)) \not \in  L_1$. Moreover, CA does not collapse the pair of arcs 
$(\pref(v), v), (v, \suff(v))$ as this would disconnect~$w$ from the rest of the solution.

\item $w$ is not closed. Denote by~$v_1$ and~$v_k$ its first and last nodes. All other nodes of~$w$ in $\operatorname{above}(L_2,2)$ are balanced. What we want to show is that in $L_1$ the only edges between levels~1 and~2 that connect~$w$ to the rest of the solution are $(\pref(v_1), v_1)$ and $(v_k, \suff(v_k))$.

We prove this for $v_k$ (for $v_1$ is it shown similarly). By Lemma~\ref{lemma:toepsilon}, there is a~path from $v_k$ to $\varepsilon$ in $L_2$ and hence $(v_k, \suff(v_k)) \in L_2$. The algorithm CA always works with an Eulerian solution and hence every node is balanced at every stage (i.e., its in-degree is equal to its out-degree). This means that $(v_k, \suff(v_k)) \in L_1$ and that all intermediate nodes of~$w$ are not connected to level~1 nodes in~$L_1$.
\end{itemize}

\item {\em Level $i=0$.} Note that $\operatorname{above}(L_1,1)$ is a~collection of walks. The case of a~non-closed walk in this case is easy as it must be connected to~$\varepsilon$ directly. For this reason, we focus on a~closed walk~$w$ in $\operatorname{above}(L_1,1)$.

We show that for every node~$v$ of~$w$ with $|v|=1$, $L_1$ contains arcs $(\varepsilon, v)$ and $(v, \varepsilon)$ (recall that for $|v|=1$, $\pref(v)=\suff(v)=\varepsilon$). This suffices as then CA (when processing level one nodes) collapses all nodes of~$w$ at level~1 except for the lexicographically largest one, and this is exactly how~$w$ is connected to~$\varepsilon$ in~$D_0$. Below, we show that $(\varepsilon, v) \in L_1$. It then follows that $(v, \varepsilon) \in L_1$ (as $L_1$ must be Eulerian).

Lemma~\ref{lemma:toepsilon} guarantees that $L_1$~contains a~path from~$v$ to~$\varepsilon$ that does not contain nodes at level~3. If the first arc of this path goes down to~$\varepsilon$, then there is nothing to prove. Hence, consider a~case when the first arc goes up to a~node~$u$ (and hence $v=\pref(u)$). The next arc then must go down to~$\suff(u)$.
Hence, $(\pref(u), u), (u, \suff(u)) \in D_1$. This may happen in two cases only: either $u$~is an~input string (i.e., $u \in \mathcal{S}$) or $u$~is the last chance of its component to be connected to the rest of the solution (i.e., exactly for this reason GHA added these two edges to the solution). The former case is straightforward: then there were at least two copies of the arcs $(\pref(u), u), (u, \suff(u))$ and CA collapsed at least one copy. Let us then focus on the latter case.

Let $x,y \in \mathcal{S}$ be such that $u=\suff(y)$ and $c := \suff(x)=\pref(y)$, see the picture below (solid arcs belong to~$L$, dashed arc belong to $L_2$).

\begin{mypic}
\node[vertex] (u) at (3,3) {$u$};
\node[vertex] (v) at (2,2) {$v$};
\node[vertex] (eps) at (1,1) {$\varepsilon$};
\node[vertex] (vv) at (0,2) {};
\node[inputvertex] (y) at (2,4) {$y$};
\node[inputvertex] (x) at (0,4) {$x$};
\node[vertex] (xl) at (-1,3) {};
\node[inputvertex] (z) at (4,4) {};
\node[vertex] (xy) at (1,3) {$c$};

\foreach \s/\t in {y/u, x/xy, xy/y, u/z, xl/x}
  \draw[->] (\s) -- (\t);
\foreach \s/\t in {xl/vv, vv/xy, xy/v, v/u}
  \draw[->,dashed] (\s) -- (\t);
\end{mypic}

Note that 
\[v = \pref(u)=\pref(\suff(y))=\suff(\pref(y))=\suff(c) \, .\]
Hence, $(c,v), (v,u) \in L_2$ (resulting from collapsing at least one pair of arcs $(c,y), (y,u) \in L$). $L_2$ also contains a~pair of arcs $(\pref(x), \suff(\pref(x))), (\suff(\pref(x)), c)$. When processing the node~$c$, CA collapses the pair of arcs $(\pref(c), c), (c, v)$ as there is an arc~$(v,u)$.
Hence, $(\varepsilon, v) \in L_1$, as required. (It may be the case that $x=y$. Then $x={\tt aaa}$, $v=\{a\}$. Then the first pair of arcs of the considered path is ${\tt a} \to {\tt aa} \to {\tt a}$ and one may just drop them.)
\end{description}

As a~final remark, note that if a~walk~$w \in \operatorname{above}(L_i,i)$ is connected to the rest of a~solution through some a~of arcs $(\pref(v), v), (u, \suff(u))$ ($v$~and~$u$ may coincide), then any other balanced node in~$w$ at level~$i$ can be fully collapsed, as every such collapse, thanks to Lemma~\ref{lem:path}, does not disconnect~$w$ or any other walk from~$\operatorname{above}(L_i,i)$ from the rest of the solution.
\end{proof}

\section{Further Directions and Open Problems}
The most immediate open problems are to prove the Collapsing Conjecture or the Weak Greedy Hierarchical Conjecture.

\subsection{Applications of Hierarchical Graphs}
It would also be interesting to find other applications of the 
hierarchical graphs. We list two such potential applications below.
\begin{description}
\item[Exact algorithms.] Can one use hierarchical graphs to solve SCS exactly in time $(2-\varepsilon)^n$?
It was shown in Section~\ref{sec:intro} that the SCS problem is a special case of the Traveling Salesman Problem. The best known exact algorithms for Traveling Salesman run in time $2^n \poly(|\inp|)$~\cite{B1962, HK1971, KGK1977, K1982, BF1996}. These algorithms stay the best known for the SCS problem as well. The hierarchical graphs were introduced~\cite{scs_exact} for an algorithm solving SCS on strings of length at most $r$ in time $(2-\varepsilon)^n$ (where $\varepsilon$ depends only on $r$). Can one use the hierarchical graph to solve exactly the general case of SCS in time $(2-\varepsilon)^n$ for a constant $\varepsilon$?

\item[Genome assembly.] The hierarchical graph in a~sense
generalizes de Bruijn graph. The latter one is heavily used
in genome assembly~\cite{pevzner2001eulerian}.
Can one adopt the hierarchical graph for this task? For this, one
would need to come up with a~compact representation of the graph
(as datasets in genome assembly are massive) as well as with a~way of
handling errors in the input data. Cazaux and Rivals~\cite{cazaux2018hierarchical} propose a linear-space counterpart of the hierarchical graph.
\end{description}

\subsection{Optimal Cycle Covers}
A superstring corresponds to a Hamiltonian path in the overlap graph, thus, a~minimum-weight cycle cover gives a~natural lower bound on its length. 
The Greedy Conjecture claims that a greedy solution never exceeds twice the length of an optimal solution. It is also  believed (see, e.g., \cite{weinard2006greedy,laube2005conditional}) that the greedy solution does not exceed the length of an optimal solution plus the length of an optimal cycle cover. 
This has interesting counterparts in the hierarchical graphs. 
\begin{itemize}
\item Note that an optimal cycle cover in the overlap graph can be constructed by a~straightforward greedy algorithm: keep taking heavy edges till the cycle cover is constructed. The proof of correctness of this algorithm relies on the Monge inequality. Interestingly, to construct an optimal cycle cover in the hierarchical graph, it suffices to invoke the Greedy Hierarchical Algorithm with lines 7--11 commented out! In a~sense, the Monge inequality is satisfied in the hierarchical graph automatically as it contains more information about input strings than just its pairwise overlaps. \item As discussed in Section~\ref{subsec:scs3}, for strings of length~3 even a~more general fact than Collapsing Conjecture holds: it suffices to have double edges adjacent to input strings. One simple way to force a~particular solution to satisfy this property is to double every edge of it. At the same time, adding a~shortest cycle cover to it is guaranteed to be as good.
\item Hence, the more general version of the Collapsing Conjecture is the following: take any solution, add any cycle cover to it, and collapse; the result is always the same. We tested this stronger conjecture and did not find any counter-examples.
\end{itemize}

%

\bibliographystyle{alpha}
\bibliography{main}

\newcommand{\etalchar}[1]{$^{#1}$}
\begin{thebibliography}{BJL{\etalchar{+}}91}

\bibitem[Bel62]{B1962}
Richard Bellman.
\newblock {Dynamic Programming Treatment of the Travelling Salesman Problem}.
\newblock {\em J. ACM}, 9:61--63, 1962.

\bibitem[BF96]{BF1996}
Eric Bax and Joel Franklin.
\newblock {A Finite-Difference Sieve to Count Paths and Cycles by Length}.
\newblock {\em Inf. Process. Lett.}, 60:171--176, 1996.

\bibitem[BJL{\etalchar{+}}91]{BJLTY1991}
Avrim Blum, Tao Jiang, Ming Li, John Tromp, and Mihalis Yannakakis.
\newblock {Linear approximation of shortest superstrings}.
\newblock In {\em STOC 1991}, pages 328--336. ACM, 1991.

\bibitem[CJR18]{cazaux2018practical}
Bastien Cazaux, Samuel Juhel, and Eric Rivals.
\newblock Practical lower and upper bounds for the shortest linear superstring.
\newblock In {\em SEA 2018}, volume 103, pages 18:1--18:14. LIPIcs, 2018.

\bibitem[CR16]{cr16}
Bastien Cazaux and Eric Rivals.
\newblock A linear time algorithm for shortest cyclic cover of strings.
\newblock {\em J. Discrete Algorithms}, 37:56--67, 2016.

\bibitem[CR18a]{cazaux2018hierarchical}
Bastien Cazaux and Eric Rivals.
\newblock Hierarchical overlap graph.
\newblock {\em arXiv preprint arXiv:1802.04632}, 2018.

\bibitem[CR18b]{cazaux20143}
Bastien Cazaux and Eric Rivals.
\newblock Relationship between superstring and compression measures: New
  insights on the greedy conjecture.
\newblock {\em Discrete Appl. Math.}, 245:59--64, 2018.

\bibitem[Gal82]{phdthesis}
John Gallant.
\newblock {\em String compression algorithms.}
\newblock PhD thesis, Princeton, 1982.

\bibitem[git18]{github}
Collapsing superstring conjecture. {G}ithub repository.
\newblock
  \url{https://github.com/alexanderskulikov/greedy-superstring-conjecture},
  2018.

\bibitem[GKM13]{GKM13}
Alexander Golovnev, Alexander~S. Kulikov, and Ivan Mihajlin.
\newblock Approximating shortest superstring problem using de {B}ruijn graphs.
\newblock In {\em CPM 2013}, pages 120--129. Springer, 2013.

\bibitem[GKM14]{scs_exact}
Alexander Golovnev, Alexander~S. Kulikov, and Ivan Mihajlin.
\newblock Solving {SCS} for bounded length strings in fewer than $2^n$ steps.
\newblock {\em Inf. Process. Lett.}, 114(8):421--425, 2014.

\bibitem[GMS80]{GMS1980}
John Gallant, David Maier, and James~A. Storer.
\newblock {On finding minimal length superstrings}.
\newblock {\em J. Comput. Syst. Sci.}, 20(1):50--58, 1980.

\bibitem[GP14]{gevezes2014shortest}
Theodoros~P. Gevezes and Leonidas~S. Pitsoulis.
\newblock {\em The shortest superstring problem}, pages 189--227.
\newblock Springer, 2014.

\bibitem[HK71]{HK1971}
Michael Held and Richard~M. Karp.
\newblock {The Traveling-Salesman Problem and Minimum Spanning Trees}.
\newblock {\em Math. Program.}, 1:6--25, 1971.

\bibitem[Kar82]{K1982}
Richard~M. Karp.
\newblock {Dynamic Programming Meets the Principle of Inclusion and Exclusion}.
\newblock {\em Oper. Res. Lett.}, 1(2):49--51, 1982.

\bibitem[KGK77]{KGK1977}
Samuel Kohn, Allan Gottlieb, and Meryle Kohn.
\newblock {A Generating Function Approach to the Traveling Salesman Problem}.
\newblock In {\em ACN 1977}, pages 294--300, 1977.

\bibitem[KS05]{KS2005}
Haim Kaplan and Nira Shafrir.
\newblock {The greedy algorithm for shortest superstrings}.
\newblock {\em Inf. Process. Lett.}, 93(1):13--17, 2005.

\bibitem[KSS15]{kulikov2015greedy}
Alexander~S. Kulikov, Sergey Savinov, and Evgeniy Sluzhaev.
\newblock Greedy conjecture for strings of length 4.
\newblock In {\em CPM 2015}, pages 307--315. Springer, 2015.

\bibitem[LW05]{laube2005conditional}
Uli Laube and Maik Weinard.
\newblock Conditional inequalities and the shortest common superstring problem.
\newblock {\em Int. J. Found. Comput. Sci.}, 16(06):1219--1230, 2005.

\bibitem[Mon81]{monge}
Gaspard Monge.
\newblock M{\'e}moire sur la th{\'e}orie des d{\'e}blais et des remblais.
\newblock {\em Histoire de l'Acad{\'e}mie Royale des Sciences de Paris}, 1781.

\bibitem[Muc07]{mucha2007tutorial}
Marcin Mucha.
\newblock A tutorial on shortest superstring approximation, 2007.

\bibitem[Muc13]{M2013}
Marcin Mucha.
\newblock {Lyndon Words and Short Superstrings}.
\newblock In {\em SODA 2013}, pages 958--972. SIAM, 2013.

\bibitem[Pal14]{P14}
Katarzyna Paluch.
\newblock Better approximation algorithms for maximum asymmetric traveling
  salesman and shortest superstring.
\newblock {\em arXiv preprint arXiv:1401.3670}, 2014.

\bibitem[PTW01]{pevzner2001eulerian}
Pavel~A. Pevzner, Haixu Tang, and Michael~S. Waterman.
\newblock An eulerian path approach to {DNA} fragment assembly.
\newblock {\em Proc. Natl. Acad. Sci. U.S.A.}, 98(17):9748--9753, 2001.

\bibitem[RBT04]{romero2004experimental}
Heidi~J. Romero, Carlos~A. Brizuela, and Andrei Tchernykh.
\newblock An experimental comparison of two approximation algorithms for the
  common superstring problem.
\newblock In {\em ENC 2004}, pages 27--34. IEEE, 2004.

\bibitem[RC18]{rivals2018superstrings}
Eric Rivals and Bastien Cazaux.
\newblock Superstrings with multiplicities.
\newblock In {\em CPM 2018}, volume 105, pages 21:1--21:16, 2018.

\bibitem[SG76]{SG1976}
Sartaj Sahni and Teofilo Gonzalez.
\newblock {P-Complete Approximation Problems}.
\newblock {\em J. ACM}, 23:555--565, 1976.

\bibitem[Sto87]{storer1987data}
James~A. Storer.
\newblock {\em Data compression: methods and theory}.
\newblock Computer Science Press, Inc., 1987.

\bibitem[STV18]{svensson2018constant}
Ola Svensson, Jakub Tarnawski, and L{\'a}szl{\'o}~A. V{\'e}gh.
\newblock A constant-factor approximation algorithm for the asymmetric
  traveling salesman problem.
\newblock In {\em STOC 2018}, pages 204--213. ACM, 2018.

\bibitem[TU88]{TU1988}
Jorma Tarhio and Esko Ukkonen.
\newblock {A greedy approximation algorithm for constructing shortest common
  superstrings}.
\newblock {\em Theor. Comput. Sci.}, 57(1):131--145, 1988.

\bibitem[Tur89]{T1989}
Jonathan~S. Turner.
\newblock {Approximation algorithms for the shortest common superstring
  problem}.
\newblock {\em Inf. Comput.}, 83(1):1--20, 1989.

\bibitem[Ukk90]{ukkonen1990linear}
Esko Ukkonen.
\newblock A linear-time algorithm for finding approximate shortest common
  superstrings.
\newblock {\em Algorithmica}, 5(1-4):313--323, 1990.

\bibitem[Wat95]{waterman1995introduction}
Michael~S. Waterman.
\newblock {\em Introduction to computational biology: maps, sequences and
  genomes}.
\newblock CRC Press, 1995.

\bibitem[web18]{webpage}
Collapsing superstring conjecture. {W}ebpage.
\newblock \url{http://compsciclub.ru/scs/}, 2018.

\bibitem[WS06]{weinard2006greedy}
Maik Weinard and Georg Schnitger.
\newblock On the greedy superstring conjecture.
\newblock {\em SIAM J. Discrete Math.}, 20(2):502--522, 2006.

\end{thebibliography}
\appendix
\section{Greedy Hierarchical Algorithm and Special Cases of SCS}
\subsection{Strings of Length~2}\label{sec:ghatwo}
Gallant et al.~\cite{GMS1980} show that SCS on strings of length $3$ is $\mathbf{NP}$-hard, but SCS on strings of length at most $2$ is solvable in polynomial time. In this section we show that GHA finds an optimal solution in this case as well. We note that the standard Greedy Algorithm does not necessarily find an optimal solution in this case. For example, if ${\cal S}=\{{\tt ab}, {\tt ba}, {\tt bb}\}$, the Greedy Algorithm may first merge {\tt ab} and {\tt ba}, which would lead to a~suboptimal solution {\tt ababb}
(recall also Figure~\ref{fig:abnormalzigzag}).

First, we can assume that all input strings from~${\cal S}$ have length exactly~$2$. Indeed, since we assume that no input string is a substring of another input string, all strings of length $1$ are unique symbols which do not appear in other strings. Take any such $s_i$ of length $1$. The optimal superstring length for ${\cal S}$ is $k$ if and only if the optimal superstring length for ${\cal S}\setminus\{s_i\}$ is $k-1$. The Greedy Hierarchical Algorithm has the same behavior: In Step~\ref{alg:gha_init}, GHA will include the arcs $(\varepsilon, s_i), (s_i, \varepsilon)$ in the solution, and it will never touch the vertex $s_i$ again (because it is balanced and connected to $\varepsilon$). Thus, $s_i$ adds $1$ to the length of the Greedy Hierarchical Superstring as well. By the same reasoning, we can assume that each string of length two is primitive, i.e., contains two distinct symbols.

When considering primitive strings ${\cal S}=\{s_1,\ldots,s_n\}$ of length exactly $2$, it is convenient to introduce the following directed graph $G=(V, E)$, where $V$ contains a vertex for every symbol which appears in strings from ${\cal S}$. The graph has $|E|=n$ arcs corresponding to $n$ input strings: for every string $s_i=ab$, there is an arc from $a$ to $b$. It is known~\cite{GMS1980} that the length of an optimal superstring in this case is $n+k$ where $k$ is the minimum number such that $E$ can be decomposed into $k$ directed paths, or, equivalently:
\begin{proposition}[\cite{GMS1980}]
Let $G$ be the graph defined above, and let $G_1=(V_1,E_1),\ldots,G_c=(V_c,E_c)$ be its weakly connected components. Then the length of an optimal superstring is
\begin{align}
\label{eq:gms}
n + \sum_{i=1}^{c} {\max\left( 1, \sum_{v \in V_i}{ \frac{ |\indegree(v) - \outdegree(v)|}{2} }\right)} \; .
\end{align}
\end{proposition}

We will now show that in this case, GHA finds an optimal solution.

\begin{lemma}
Let ${\cal S}=\{s_1,\ldots,s_n\}$ be a set of strings of length at most $2$, and let $s$ be an optimal superstring for ${\cal S}$. Then $GHA({\cal S})$ returns a~superstring of length~$|s|$. 
\end{lemma}

\begin{proof}
We showed above that it suffices to consider the case of $n$ primitive strings $\{s_1,\ldots,s_n\}$ of length exactly $2$. For $1\leq i\leq n$, let $s_i=a_i b_i$, where $a_i\neq b_i$. Consider the partial greedy hierarchical solution $D$ after the Step~\ref{alg:gha_init} of the GHA algorithm: $D=\{(a_i, a_ib_i), (a_ib_i, b_i): 1\leq i\leq n \}$. (We abuse notation by identifying the set of arcs $D$ with the graph induced by $D$.) This partial solution has $n$ up-arcs, so its current weight is $n$. 

Note that by the definition of the graph $G$ above, $G$ contains an arc $(a, b)$ if and only if $D$ has the arcs $(a, ab)$ and $(ab, b)$ of the graph HG. Thus, the indegree (outdegree) of a vertex $a$ in $G$ equals the indegree (outdegree) of the vertex $a$ in the partial solution $D$. Also, two vertices $a$ and $b$ of $G$ belong to one weakly connected component in $G$ if and only if they belong to one weakly connected component in $D$. Therefore, the expression~\eqref{eq:gms} in $G$ has the same value in the partial solution graph $D$. (Indeed, the vertices of $D$ corresponding to strings of length $2$ are balanced and do not form weakly connected components.) 

Now we proceed to Steps~\ref{alg:for}--\ref{alg:last} of GHA. GHA will go through all strings of length $1$, and add $|\indegree(v) - \outdegree(v)|$ arcs for each unbalanced vertex $v$. The Steps~\ref{alg:else}--\ref{alg:last} ensure that each weakly connected component adds at least a pair of arcs. Since exactly a half of added arcs are up-arcs, we have increased the weight of the partial solution $D$ by
\begin{align*}
\sum_{i=1}^{c} {\max\left( 1, \sum_{v \in V_i}{ \frac{ |\indegree(v) - \outdegree(v)|}{2} }\right)} \; .
\end{align*}
\end{proof}

\subsection{Spectrum of a~String}\label{sec:ghaspectrum}
By a~$k$-spectrum of a~string $s$ 
(of length at least~$k$)
we mean a~set of all substrings of~$s$ of length~$k$.
Pevzner et al.~\cite{pevzner2001eulerian} give a polynomial time exact algorithm for the case when the input strings form a spectrum of an unknown string. We show that GHA also finds an optimal solution in this case.

\begin{lemma}
Let ${\cal S}=\{s_1,\ldots,s_n\}$ be a~$k$-spectrum of an unknown string~$s$. Then $GHA({\cal S})$ returns a~superstring of length at most~$|s|$.
\end{lemma}

\begin{proof}
Since $s$ has $n$ distinct substrings of length $k, \, |s|\geq n+k-1$. We will show that GHA finds a superstring of length $n+k-1$. After Step~\ref{alg:gha_init} of GHA, the partial solution $D = \{(\operatorname{pref}(s), s), (s, \operatorname{suff}(s)) \colon s \in {\cal S}\}$. In particular, $D$ is of weight $n$. For $1\leq i\leq k-1$, let $u_{i}$ be the first $i$ symbols of $s$, and let $v_{i}$ be the last $i$ symbols of $s$. Note that $u_{k-1}$ and $v_{k-1}$ are the only unbalanced vertices of the partial solution $D$ after Step~\ref{alg:gha_init}: all other strings of length $k-1$ appear equal number of times as prefixes and suffixes of strings from ${\cal S}$. Therefore, while processing the level $\ell=k-1$, GHA will add one arc to each of the vertices $u_{k-1}$ and $v_{k-1}$, and will not add arcs to other strings of length $k-1$. 

In general, while processing the level $\ell=i$, GHA adds one up-arc to $u_i$ and one down-arc to $v_i$. In order to show this, we consider two cases. If $u_i \neq v_i$, then $u_i$ has an incoming arc from the previous step and does not have outgoing arcs, therefore GHA adds an up-arc to $u_i$ in Step~\ref{alg:step6}. Similarly, GHA adds a down-arc from $v_i$. Note that there are no other strings of length $i<k-1$ in the partial solution, so the algorithm moves to the next level. In the case when $u_i=v_i$, we have that all vertices are balanced, but the string $u_i$ is now the shortest string in this only connected component ${\cal C}$ of the graph. Therefore, for $i>0$ we have $\varepsilon\not\in{\cal C}$, and GHA adds an up- and down-arc to $u_i$ in Step~\ref{alg:last}. 

We just showed that GHA solution for a $k$-spectrum of a string has the initial set of arcs  $D = \{(\operatorname{pref}(s), s), (s, \operatorname{suff}(s)) \colon s \in {\cal S}\}$, and also the arcs $\{ (u_{i-1}, u_{i}), (v_i, v_{i-1})\colon 1\leq i\leq k-1 \}$. Thus, the total number of up-arcs (and the weight of the solution) is $n+k-1$.
\end{proof}

\subsection{Tough Dataset}\label{sec:ghatough}
There is a~well-known dataset consisting of just three strings where the classical greedy algorithm produces a~superstring that is almost twice longer than an optimal one: $s_1={\tt cc}({\tt ae})^n$, $s_2=({\tt ea})^{n+1}$, $s_3=({\tt ae})^n{\tt cc}$. Since $\overlap(s_1, s_3)=2n$,
 while $\overlap(s_1,s_2)=\overlap(s_2,s_3)=2n-1$, the greedy algorithm produces a~permutation $(s_1, s_3, s_2)$ (or $(s_2,s_1,s_3)$). I.e., by greedily taking the massive overlap of length $2n$ it loses the possibility to insert $s_2$ between $s_1$ and $s_3$ and to get two overlaps of size $2n-1$. The resulting superstring has length $4n+6$. At the same time, the optimal superstring corresponds to the permutation $(s_1,s_2,s_3)$ and has length $2n+8$.
 
The algorithm GHA makes a~similar mistake on this dataset, see Figure~\ref{fig:tough}. When processing the node $({\tt ea})^n$, GHA does not add two lower arcs to it and misses a~chance to connect two components. It is then forced to connect these two components through~$\varepsilon$. This example shows that GHA also does not give a better than $2$-approximation for SCS.

%
%
%
%

\newcommand{\ged}[2]{
\tikzstyle{t}=[vertex,draw=black]
\tikzstyle{p}=[->,decorate,decoration=snake,bend left=10]
\begin{scope}[yshift=#1mm]
\node[inputvertex] (ccaen) at (0,5) {{\tt cc}({\tt ae})$^n$};
\node[inputvertex] (eann) at (10,5) {({\tt ea})$^{n+1}$};
\node[inputvertex] (aencc) at (4.5,5) {({\tt ae})$^n${\tt cc}};
\node[t] (ccaena) at (-1,4) {\tt cc(ae)$^{n-1}$a};
\node[t] (caen) at (1.5,4) {\tt c(ae)$^{n}$};
\node[t] (eane) at (9,4) {\tt (ea)$^{n}$e};
\node[t] (aean) at (11,4) {\tt a(ea)$^{n}$};
\node[t] (aenc) at (3.5,4) {\tt (ae)$^{n}$c};
\node[t] (eaencc) at (6,4) {\tt e(ae)$^{n-1}$cc};
\node[t] (ccaenn) at (1,3) {\tt cc(ae)$^{n-1}$};
\node[t] (aen) at (3.5,3) {\tt (ae)$^{n}$};
\node[t] (ean) at (10,3) {\tt (ea)$^{n}$};
\node[t] (aenncc) at (6,3) {\tt (ae)$^{n-1}$cc};
\node[t] (eaenn) at (3.5,2) {\tt e(ae)$^{n-1}$};
\node[t] (aeann) at (7,2) {\tt a(ea)$^{n-1}$};
\node[t] (eps) at (5,0) {$\varepsilon$};

\foreach \f/\t in {ccaena/ccaen, ccaen/caen, eane/eann, eann/aean, aenc/aencc, aencc/eaencc, ccaenn/ccaena, caen/aen, ean/eane, aean/ean, aen/aenc, eaencc/aenncc, ean/aeann, eaenn/ean}
  \draw[->] (\f) -- (\t);
  
\path (eps) edge[p] (ccaenn);
\path (aenncc) edge[p] (eps);

#2
\end{scope}
}

\begin{figure}[!ht]
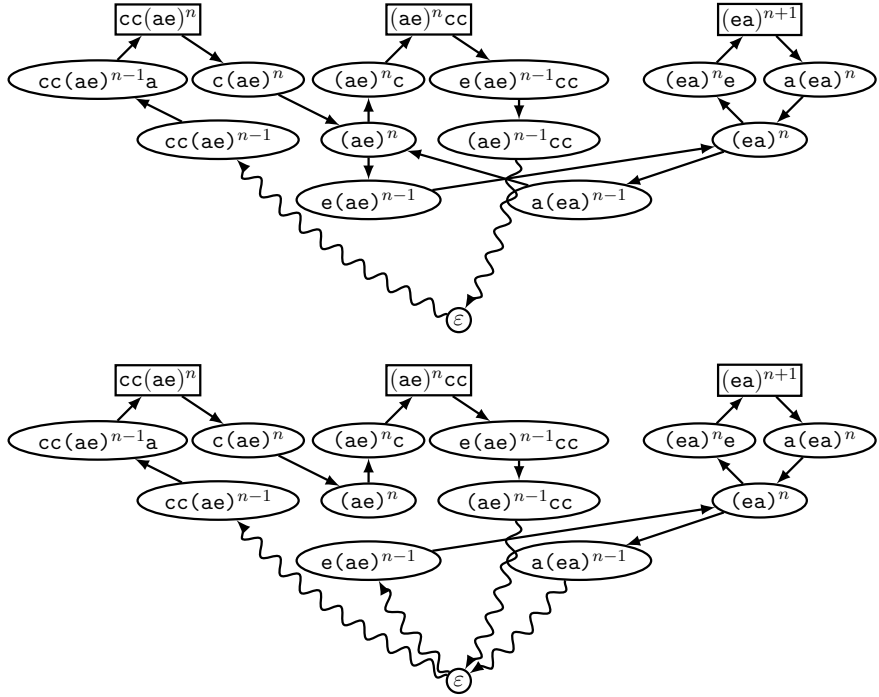

\begin{mypic}
\ged{0}{
\foreach \f/\t in {aeann/aen, aen/eaenn}
  \draw[->] (\f) -- (\t);
}

\ged{-60}{
\path (eps) edge[p] (eaenn);
\path (aeann) edge[p] (eps);
}
\end{mypic}
\caption{Top: optimal solution for the dataset $\{ {\tt cc}({\tt ae})^n, ({\tt ea})^{n+1}, ({\tt ae})^n{\tt cc} \}$. Bottom: solution constructed by GHA.}
\label{fig:tough}
\end{figure}

\end{document}